\documentclass[sigplan,10pt]{acmart}
\renewcommand\footnotetextcopyrightpermission[1]{}
\usepackage[english]{babel}

\usepackage{tikz}
\usepackage{filecontents}

\usepackage{tikz}
\usepackage{xcolor}
\usepackage{xspace}
\usepackage{tikz}
\usepackage{pifont}
\usepackage[english]{babel}
\usepackage{blindtext}
\usepackage{lipsum}

\usepackage[ruled,vlined,linesnumbered]{algorithm2e}
\usepackage{booktabs}

\usepackage{filecontents}
\usepackage{xspace}
\usepackage{xcolor}
\usepackage{hhline}
\usepackage{multirow}
\usepackage{soul}

\usepackage{array}
\usepackage{comment}
\usepackage{listings}
\usepackage{dsfont}

\usepackage{algorithmic}
\usepackage{xurl}
\usepackage{setspace}
\usepackage[flushleft]{threeparttable}
\usepackage{enumerate}
\usepackage{float}
\usepackage{subcaption}
\usepackage{graphicx}
\usepackage{breqn}
\usepackage{enumitem}

\usepackage{caption}
\usepackage{ltablex}

\usepackage{ifluatex}

\usepackage{outlines}

\usepackage{comment}
\usepackage{amsmath, amssymb}
\definecolor{darkkhaki}{rgb}{0.74, 0.72, 0.42}

\newcommand{\name}{Argo\xspace}

\newcommand{\anonemail}{\texttt{anon\_mail}\xspace}

\newcounter{packednmbr}
\newenvironment{packedenumerate}{\begin{list}{\thepackednmbr.}{\usecounter{packednmbr}\setlength{\itemsep}{0.5pt}\addtolength{\labelwidth}{-4pt}\setlength{\leftmargin}{2ex}\setlength{\listparindent}{\parindent}\setlength{\parsep}{1pt}\setlength{\topsep}{0pt}}}{\end{list}}
\newenvironment{packeditemize}{\begin{list}{$\bullet$}{\setlength{\itemsep}{0.5pt}\addtolength{\labelwidth}{-4pt}\setlength{\leftmargin}{2ex}\setlength{\listparindent}{\parindent}\setlength{\parsep}{1pt}\setlength{\topsep}{2pt}}}{\end{list}}

\newcommand{\tightcaption}[1]{\vspace{-0.4cm}\caption{{\normalfont{{{#1}}}}}\vspace{-0.3cm}}
\newcommand{\tightsection}[1]{\vspace{-0.2cm}\section{#1}\vspace{-0.2cm}}

\newcommand{\eg}{{\it e.g.,}\xspace}

\newcommand{\mypara}[1]{\vspace{0.05cm}\noindent{\bf {#1}:}~}

\newcommand{\myparaq}[1]{\smallskip\noindent{\bf {#1}?}~}

\definecolor{backcolour}{rgb}{0.96,0.96,0.96}
\definecolor{codegray}{rgb}{0.5,0.5,0.5}
\definecolor{deepblue}{rgb}{0,0,0.6}
\definecolor{deepred}{rgb}{0.6,0,0}
\definecolor{deepgreen}{rgb}{0,0.5,0}
\lstdefinestyle{mystyle}{
    backgroundcolor=\color{backcolour},   
    commentstyle=\color{codegreen},
    morekeywords={self, True},
    keywordstyle=\color{deepblue},
    numberstyle=\tiny\color{codegray},
    emph={MyClass,__init__,EncodingType,Image},
    emphstyle=\color{deepred},
    stringstyle=\color{deepgreen},
    basicstyle=\ttfamily\footnotesize,
    breakatwhitespace=false,         
    breaklines=true,                 
    captionpos=b,                    
    keepspaces=true,                 
    numbers=left,                    
    numbersep=5pt,                  
    showspaces=false,                
    showstringspaces=false,
    showtabs=false,                  
    tabsize=1
}

\usepackage{tcolorbox}
\tcbuselibrary{listings, skins, breakable, xparse, theorems}
\usepackage{xfrac}

\begin{document}

\newcommand*{\affmark}[1][*]{\textsuperscript{#1}}
\newcommand*{\affaddr}[1]{#1}

% Change title for ArXiv
\title{\name: Efficient Importance Labeling for Enterprise Email Systems}
\author{Siddhant Ray}
\affiliation{
  \institution{University of Chicago}
  \city{}
  \country{}
}

\author{Ganesh Ananthanarayanan}
\affiliation{
  \institution{Microsoft}
  \city{}
  \country{}
}

\author{Kevin Chian}
\affiliation{
  \institution{Microsoft}
  \city{}
  \country{}
}

\author{Yan Guo}
\affiliation{
  \institution{Microsoft}
  \city{}
  \country{}
}

\author{Cristina St Hill}
\affiliation{
  \institution{Microsoft}
  \city{}
  \country{}
}

\author{Jack W. Stokes}
\affiliation{
  \institution{Microsoft}
  \city{}
  \country{}
}

\author{Victor Wang}
\affiliation{
  \institution{Microsoft}
  \city{}
  \country{}
}

\author{Junchen Jiang}
\affiliation{
  \institution{University of Chicago / TensorMesh }
  \city{}
  \country{}
}
\renewcommand{\shortauthors}{}

\begin{abstract}
Email importance labeling has long been a critical yet challenging problem for businesses and individuals. Traditional approaches—such as keyword matching, user-defined rules, and sender-based heuristics—demand extensive manual feature engineering and fail to scale effectively or generalize. Recent advances in large language models (LLMs) demonstrate strong potential and a natural fit for this task, offering deep contextual understanding and superior labeling quality. However, using LLM models like GPT-4.1 at enterprise email volumes incurs prohibitive computational costs and hinders real-world deployment. We explore the trade-off space of using alternative labeling schemes as opposed to GPT4.1 scale LLMs, with the goal of achieving near-GPT-level labeling quality with significantly lower cost. We develop \name, an enterprise-email labeling framework, where we construct a profiler to efficiently search the cost-quality tradeoff-space of labeling and identify cost-efficient alternatives to labeling emails. Additionally, we design an on-demand provisioning scheme to intelligently scale \name with real-time load, to minimize cost-increases during peak-load inference. Over 3 open-source email datasets, \name achieves 148-167$\times$ inference cost reduction with negligible quality degradation and 20-640000$\times$ lower profiling costs, making large-scale,context-aware email labeling practical for enterprises.
\end{abstract}

\maketitle
\pagestyle{plain}
\vspace{-0.78em}

% \pagestyle{plain}

%-------------------------------------------------------------------------------
\section{Introduction}
\label{sec:intro}

Email importance labeling streamlines both personal and organizational email workflows by directing user attention toward urgent tasks and time-sensitive messages while pushing low-value correspondence out of the critical path~\cite{Klimt2004, Bekkerman2004}. It improves continuity by making important messages and threads easier to track and revisit, and enables cleaner task management by organizing action items and related data into appropriate folders~\cite{Ducheneaut2001}. Reliable and accurate email importance labeling on user inboxes boosts employee productivity, improves workflows and enables efficient time management for the employees and the businesses.

Email labeling is a long-standing, well-studied problem, with early systems relying heavily on manually engineered features to approximate message importance ~\cite{Klimt2004, Bekkerman2004}. Classical schemes are fundamentally limited: simple keyword-matching approaches (\eg detecting “urgent”) lack contextual understanding and produce low-quality labels~\cite{whittaker}. Systems using sender-side engineered features (\eg sender importance) or email properties (\eg CC-list structure or message length)~\cite{Ducheneaut2001, 10.1145/3397271.3401121}, provide stronger but incomplete signals. These heterogeneous signals demand extensive manual feature selection, which does not scale across large, combinatorial feature subsets. This makes learning email importance labeling inherently limited by the underlying feature engineering complexities~\cite{Wang2019, Yang2017, Alkhereyf2017}. Earlier works also heavily focused on spam filtering, which removed irrelevant emails and reduced inbox clutter~\cite{modardi,whittaker}, importance labeling for emails is complementary, focusing on boosting productivity.

We revisit enterprise-level email-importance labeling in the era of highly expressive large language models (LLMs). LLMs are a natural fit for this task: state-of-the-art language modeling provides deep contextual understanding, enabling substantially higher-quality labeling than classical feature-engineered systems. Modern LLMs are trained extensively on large text corpora for text generation and reasoning tasks, making them particularly well-suited for enterprise-level email labeling. As email content is almost entirely natural language (\eg limited metadata such as timestamps are not), LLMs can operate directly on the raw input without requiring manual featurization or combinatorial feature selection. This eliminates the need to engineer sender-signals, structural properties, or keyword heuristics, and allows the model to infer relevance and intent from the semantic context.

A reliable and well-integrated labeling feature strengthens an enterprise-email client’s competitiveness, increases user adoption and retention and ultimately supports revenue growth~\cite{Mackay1988}. Given this, leading email clients are looking to adopt such labeling solutions and the industry trends already showcase labeling methods for similar applications (\eg Apple's iMessage~\cite{apple2025summarize_notifications}). We carry out this work with a leading email service deploying such solutions in production, who we will refer to as \anonemail for the remaining paper.

\begin{figure}[t]
    \centering
    \includegraphics[width=\linewidth]{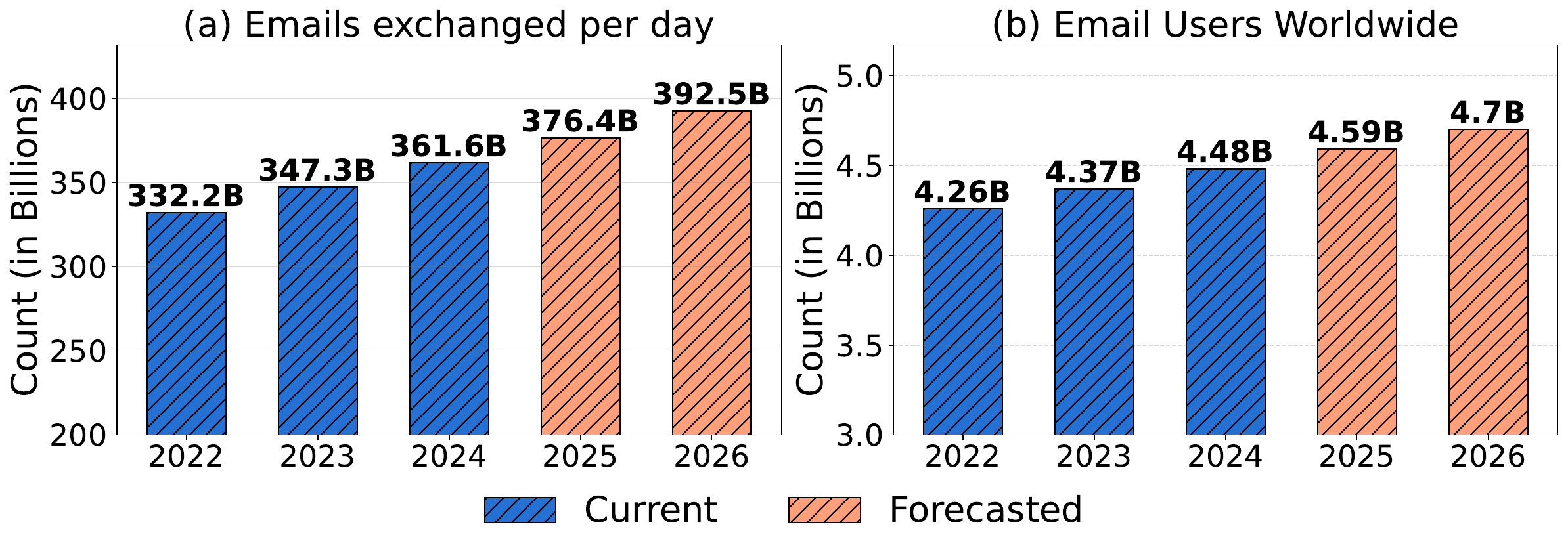}
    \vspace{-0.2cm}
    \tightcaption{At real-scale and increasing email exchange volumes, cost of labeling with LLMs becomes \emph{prohibitive}}
    \label{fig:intro}
\end{figure}

However, enterprise-level email workloads have extremely high volumes. Figure \ref{fig:intro} (a) shows the number of emails being exchanged per day across the world and Figure \ref{fig:intro} (b) shows the increasing number of email users over the last decade. While LLMs like GPT4.1\cite{openai2024gpt4technicalreport} offer superior labeling quality compared to classical methods, running labeling with an LLM at this scale leads to \emph{prohibitive running costs} and resource utilization, making it infeasible to naively use it for generating every email label. Leading email providers like \anonemail, while desiring such labeling quality, are balking at the cost of labeling at this scale (\eg GPT4.1 labeling leads to labeling costs of 5-10 billion USD per month at current email volumes). Given this, we ask the question, how do we build an \emph{efficient system} which achieves the best quality-cost tradeoffs for email labeling by intelligently balancing quality and cost? Further, as real-time email arrival loads vary significantly over time, we need \emph{efficient} resource provisioning to allow deployed enterprise-level email to label efficiently at peak load, without incurring significantly increased costs.

We have adapted lower-cost alternative systems to LLMs to use for email labeling, using either cascades of Small Language Models (SLMs)~\cite{soiffer-etal-2025-semantic, ICLR2024_11f5520d, lu-etal-2025-demystifying}  or trained classifiers based text embeddings~\cite{conneau-etal-2017-supervised, reimers-gurevych-2019-sentence, wang-etal-2024-improving-text} (more in \S \ref{ssec:SLM_only} and \ref{ssec:embedding_only}). However, these systems have \emph{multiple knobs} (\eg which SLMs to use, different labeling requirements etc.) which leads to an \emph{extremely large search space} (> 100M configurations, more in \S \ref{ssec:large_space})  over heterogeneous email sets. To utilize such systems for achieving best cost-quality tradeoffs for labeling, we need to design an \emph{efficient} profiler to navigate the search space and manage inference cost. We introduce \name, which has (a) an efficient profiler to select labeling methods achieving best quality-cost tradeoffs and identify a \emph{Balanced} point for optimal tradeoff and (b) a resource-provisioning scheme to minimize cost increases under peak email load.

Unlike several traditional profiler-based systems, messaging workloads like emails, do not have clear spatial or temporal insights to be utilized for profiling (\eg in contrast, several video analytics profiling systems~\cite{focus_osdi2018, chameleon_sigcomm2018, videostorm_nsdi17, madeye_nsdi2024} relied on such properties). \name identifies a novel insight, the 
characteristics and needs of the email labels can be used as a signal to profile and choose the labeling method. These characteristics provide clear input groupings and allow us to make offline-profiled static choices for parameters at runtime. For example, the complexity of labels (\eg binary vs non-binary) allows \name to decide the choice of the labeling method for assigning its final value (more in \S \ref{sec:motivation} and \ref{sec:design}).

\name allows a dynamic choice between both a cascade built of Small Language Models (SLMs) with increasing expressiveness and a lightweight embedding-trained classifier, to assign labels to the emails based on the characteristics of the label. Offline profiling on a calibration set of emails allows \name to decide the appropriate label-method match. Additionally, \name's profiler provides all necessary details on the choice of SLMs in the cascade, the confidence thresholds used to determine the correct point of operation in the cascade etc. \name also uses central properties (\eg error tolerance, asynchronous nature) of the cost-quality tradeoff space for email labeling to make more detailed labeling decisions. 

Additionally, running email labeling at enterprise scale requires an efficient system to manage resources to match the evolving arrival of emails. Services like \anonemail are provisioned to use SLMs via \emph{model-as-service} at a certain  \emph{averaged} capacity and are charged a penalty for exceeding it, which often occurs during peak load (\S \ref{ssec:provisioning}). \name develops a greedy resource provisioning algorithm for the SLM cascade, to balance between further provisioning SLM instances which have reached capacity and using other under-capacity SLMs in the cascade, to minimize cost increases during load.

We make the following contributions, incorporating the needs of \anonemail looking to adopt cost-efficient email labeling in production -

\begin{itemize}
    \item \name is the first system for \emph{cost-efficient} labeling of email enterprise workloads with near LLM labeling accuracy while reducing inference costs by 148-167$\times$.
    \item \name efficiently searches the quality-cost tradeoff space for email labeling and chooses a \emph{Balanced} point with 20-640000$\times$ lower profiling costs with no quality drop.
    \item \name provides a resource provisioning scheme to minimize cost-increase of labeling under increased load by 2.2-3.8$\times$.
\end{itemize}
% \section{Motivation and Related Work}
\section{Background: Email Importance Labeling}
\label{sec:background}

\mypara{Email labeling schemes over time}
Email importance classification is a central problem in intelligent personal and business communication management due to increasing volumes of digital correspondence over the recent decades. Early research modeled it supervised text-classification, emphasizing manually engineered structural and behavioral features to estimate message priority~\cite{Klimt2004, Bekkerman2004, Kiritchenko2001, Lampert2010}. These feature-based approaches use both email and metadata including sender–recipient patterns, conversational structure, and temporal dynamics. Yet, they are inherently constrained by limited feature expressiveness and poor generalization across heterogeneous inputs. The introduction of large publicly available corpora such as the Enron dataset further revealed the complexity of the task, as it shows substantial variability in communication norms, organizational hierarchies, and user-specific preferences~\cite{Klimt2004, Shetty2004}.

Recent work has increasingly leveraged deep neural architectures and relational modeling to capture the contextual and semantic dimensions of email communication. Neural methods, including hierarchical attention networks and pretrained language models, have shown substantial improvements by modeling semantic structure and user-dependent patterns in a unified representation space~\cite{Deng2018, Zhang2021, Devlin2019, Yang2016}. In parallel, graph-based approaches explicitly encode communication networks, representing individuals, threads, and interactions as nodes and edges to infer message importance from structural and social signals~\cite{Diehl2019, McCallum2007, Tang2015}. 

Email labeling is also being deployed in leading email providers. Gmail provides algorithmic email-prioritization through its Priority Inbox, by ranking messages and separating them into sections such as Important \& Unread, Starred, and Everything Else. It also offers Category Tabs (Primary, Social, Promotions, Updates, Forums) that automatically classify incoming messages \cite{gmailpriority, gmailcategories}. Outlook uses its Focused Inbox system, which relies on behavioral signals to divide messages into a Focused tab for high-relevance mail and an Other tab for lower-priority items~\cite{outlookfocused}. Apple Mail supports sender-based prioritization through VIP senders and, in newer versions, provides a Priority Messages view that highlights time-sensitive or important communications~\cite{applevip}.

Despite these advances, email labeling continues to face challenges in scaling use modern LLMs to large and heterogeneous email volumes, as computational costs associated with these architectures skyrocket trying to preserve high labeling accuracy~\cite{Strubell2019, Narayanan2021}. These issues underscore the need for carefully balancing labeling quality with efficiency, enabling practical and cost-effective deployment of email-importance labeling systems in real-world organizations.

\mypara{Quality and Cost Metrics}
We consider the following metrics for evaluating labeling quality and cost.

\begin{packeditemize}
    \item \emph{F1-score} - The F1-score is the harmonic mean of precision and recall. It balances the trade-off between the two, especially useful when classes are imbalanced. We use this metric to compare the quality against the baseline LLM.
    \item \emph{API cost reduction factor} - We measure the cost reduction factor of different models and combinations with respect to the cost of using the baseline LLM for the given inputs.
\end{packeditemize}
\vspace{-0.3cm}

\section{Towards obtaining better cost-quality tradeoffs for email labeling}
\label{sec:motivation}

\subsection{How are different email labels distributed?}

Overall importance labeling for emails can have multiple individual labels with different specificity and characteristics, leading to needing different methods being optimal for assigning them. Further, inherent structures (\eg binary vs non-binary values) in the distribution and nature of such labels can further help making efficient labeling decisions, to reduce the overall labeling cost. We conduct a measurement study on the distribution of several importance labels , which can be assigned to emails in a user's inbox.

\begin{figure}[t]
    \centering
    \includegraphics[width=\linewidth]{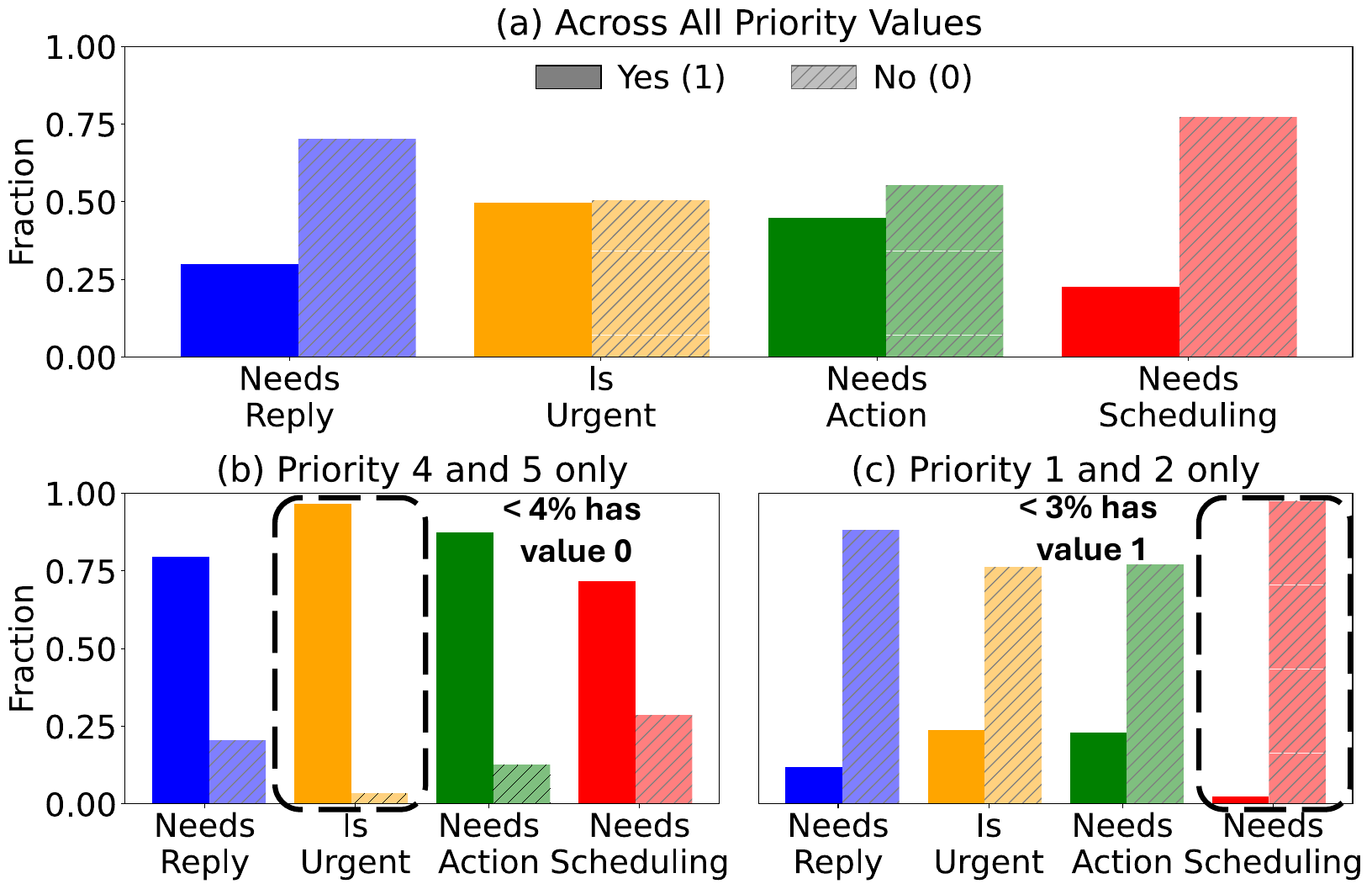}
    \vspace{-0.2cm}
    \tightcaption{Label overlaps vary significantly across different subsets which should be used to guide the labeling system}
    \vspace{-0.2cm}
    \label{fig:combined}
\end{figure}

\begin{packeditemize}
    \item \emph{Priority} - This label provides an overall importance ranking on how important the email is measured on a discrete scale from 1 (lowest) to 5 (highest).
    \item \emph{Needs Reply} - This binary label determines if an email requires a reply or not, based on either an explicit request or an implicit signal in the email text.
    \item \emph{Is Urgent} - This binary label measures the time-sensitivity and how quickly should the email be read by the receiver.
    \item \emph{Needs Action} - This binary label decides if the receiver needs to carry out an explicit task (\eg approval, admin etc.) for the person sending the email.
    \item \emph{Needs Scheduling} - This binary label determines if the email requires a follow up meeting or chat event, which needs to be scheduled in a time-sensitive manner.
\end{packeditemize}

We observe that email labeling can be be both non-binary or binary decisions. In all the labels which are binary, 1 is defined as a yes and 0 is defined as a no. The labels considered in the study are based on the business requirements of \anonemail deployed in production. While the study is guided by their needs, it does not pose any constraints in our design. The list of labels can be extended to contain more signals, based on specific client or product feedback.

We consider the well-studied Enron email dataset~\cite{Klimt2004} for the purpose of the measurement study on label distributions. The Enron dataset has 137 unique users who held positions in the senior management staff at Enron Inc. Every user has a dedicated inbox of emails, from where we draw a sample of emails for our study. Overall, we draw a $20\%$ random sample (8000 emails) from the overall set of 40000 emails across all users. As the dataset doesn't have assigned labels, we assign values to each of the 5 labels introduced above with GPT4.1.

In Figure \ref{fig:combined}, we measure the distribution overlap of the different labels and observe that not all label combinations are statistically significant. Figure \ref{fig:combined} (a), we see that if we don't set any conditions on the \emph{Priority} label and look at the distribution of other labels, we do not observe any clear patterns in the distribution. However, if we fix the \emph{Priority} label as high (\eg 4) in \ref{fig:combined} (b), we observe that we have almost no cases (less than $4\%$) where the label has a value 0 for the \emph{Is Urgent}. Similarly, if we set the \emph{Priority} label as low (\eg 2), we observe that we have almost no cases (less than $3\%$) where the label has a value 1 for the \emph{Needs Scheduling}. 

Empirically we observe that email labels have specific characteristics and not all combinations occur significantly enough to be considered jointly. This drastically reduces the combinatorial space for building the labeling scheme over the labeling combination and the system should use such insights to reduce the labeling costs. Concretely, if we run the labeling sequentially for various labels, based on the assigned value for a given label, the system can decide to not run subsequent labels for the particular input email.

\begin{figure}[t]
    \centering
    \includegraphics[width=\linewidth]{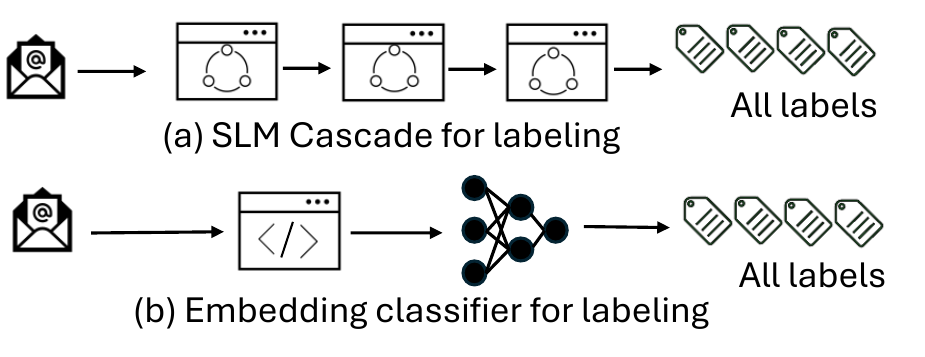}
    \vspace{-0.8cm}
    \caption{Lower-cost alternatives to LLM labeling}
    \label{fig:comb_methods}
\end{figure}

% \begin{figure}[t]
%     \centering
%     \includegraphics[width=\linewidth]{measurement_figures/cascade.pdf}
%     \tightcaption{SLM cascade for labeling}
%     \label{fig:cascade_illus}
% \end{figure}

% \begin{figure}[t]
%     \centering
%     \includegraphics[width=\linewidth]{measurement_figures/embedding.pdf}
%     \tightcaption{Embedding classifier for labeling}
%     \label{fig:embedding_illus}
% \end{figure}

\begin{table}[t]
\centering
\begin{tabular}{cc}
\hline
\textbf{Knob} & \textbf{Description} \\
\hline
$\mathcal{L}$ & SLM cascade or embedding classifier \\
$\mathcal{M}$ & Different SLMs and embedding models \\
$\mathcal{T}$ & Confidence values used in cascade \\
$\mathcal{O}$ & Order of SLMs in cascade \\
$\mathcal{S}$ & Size of calibration email set \\
\hline
\end{tabular}
\caption{Knobs to be profiled for choosing optimal tradeoffs}
\vspace{-0.6cm}
\label{tab:knobs}
\end{table}

\subsection{What if we only use an SLM cascade?}
\label{ssec:SLM_only}

Several works have proposed using SLM cascade-based solutions have for general text classification and generation tasks in the recent times~\cite{soiffer-etal-2025-semantic, ICLR2024_11f5520d, lu-etal-2025-demystifying} Figure  \ref{fig:comb_methods} (a) illustrates a pure SLM-based approach for email labeling. For generating a label value, the email is given to the first model in the cascade which generates an output and a confidence score. Based on a confidence threshold, the system decides whether to use the label or move up the cascade. This process repeats till a model generates a label with confidence above the threshold.

Using only SLM cascades for generating every label for each email does not scale. While SLM cascades can provide high labeling quality for all email labels due to having a suite of models with different capabilities, the cost of labeling remains comparable to baseline LLMs. For maintaining quality, inference needs to be run independently for each label as performing joint inference across the labels leads to a $15 - 20\%$ drop in F1-score. This is attributed to increasing complexity of jointly labeling multiple values, which often overwhelms the reasoning capability of the SLMs in the cascade. 

Hence, this requires running the cascade independently for every label, which reduces re-usability and increases the inference cost. Additionally, there's no single \emph{works-for-all} cascade. SLM cascades need \emph{multiple knobs} for operation (\eg which models in the cascade, choice of threshold), which need to be calibrated based on the workload characteristics.

\begin{figure}[t]
    \centering
    \includegraphics[width=0.95\linewidth]{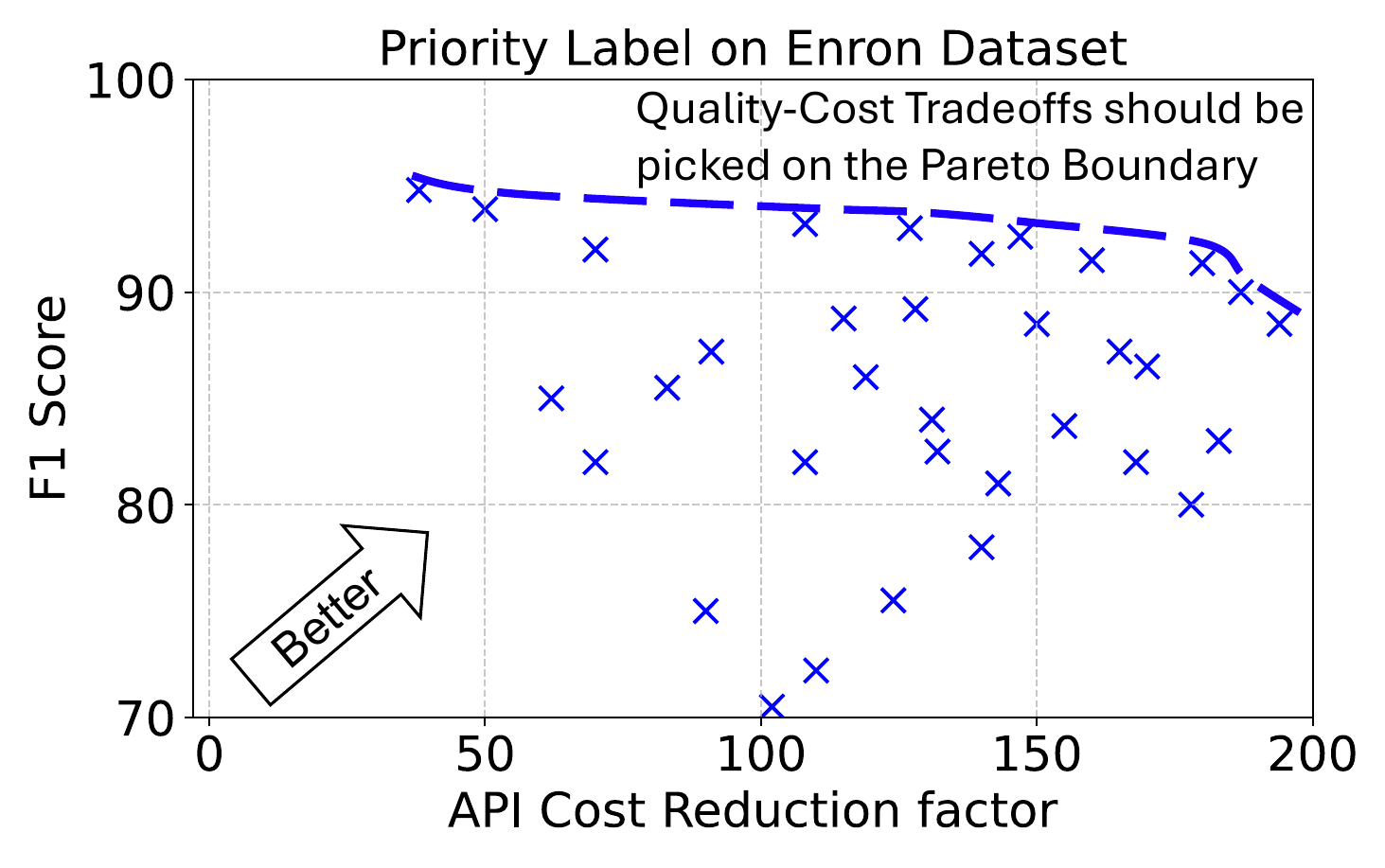}
    \tightcaption{Varying the confidence thresholds for the SLM Cascade while keeping other knobs fixed}
    \vspace{-0.42cm}
    \label{fig:comb_cost}
\end{figure}

\subsection{What if we only use an embedding classifier?}
\label{ssec:embedding_only}

Another direction of work has proposed training a neural network for text classification using semantically rich, frozen text embedding vectors~\cite{conneau-etal-2017-supervised, reimers-gurevych-2019-sentence, wang-etal-2024-improving-text}. Figure \ref{fig:comb_methods} (b) illustrates this approach. A small feed-forward neural network is trained using embeddings generated from a set of input emails, which is then used to assign label values for new emails at runtime.

Using a pure embedding classifier based solution significantly reduces labeling costs as the classifier can be jointly trained across all labels and the embeddings of the emails may be reused across different labels. Hence, for inference, each email needs to be converted into vector embedding only once. However, training such a classifier is challenging, especially for non-binary labels, due to extreme class imbalances~\cite{10.1109/TKDE.2008.239} in the training datasets and significant heterogeneity across emails. While the method yields high quality labels for labels with binary values, we see a significant drop ($> 15\% $) in F1-score in labeling values for non-binary labels.

Hence, we need to build a \emph{smart system} which \emph{profiles} the requirements of the labels to get the optimal methods for labeling, which in turn would lead to overall \emph{best quality-cost tradeoff} for labeling enterprise-scale email workloads.

\begin{figure*}[t]
    \centering
    \includegraphics[width=0.95\linewidth]{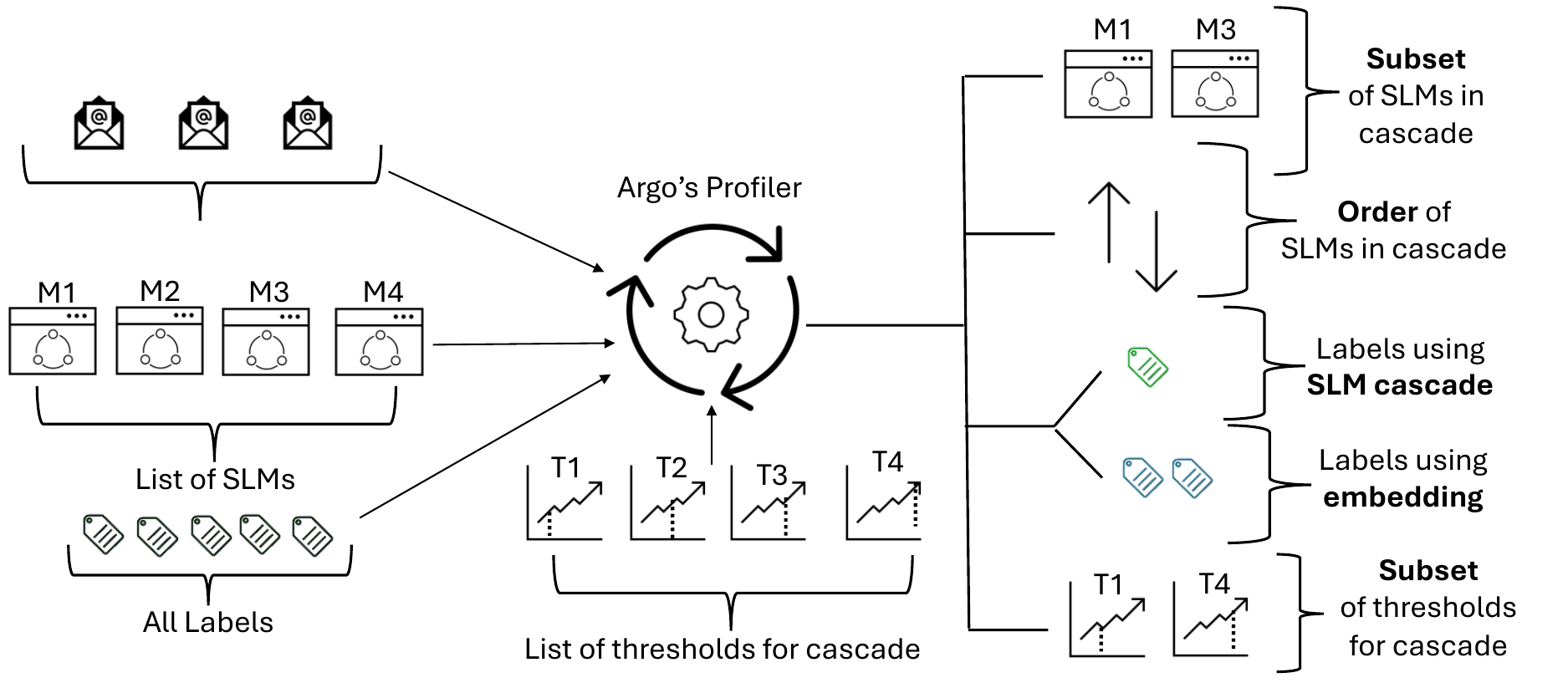}
    \tightcaption{\name performs offline profiling to determine knob values and decisions to be used for email labeling}
    \label{fig:profiler}
\end{figure*}

\subsection{Obtaining optimal quality-cost tradeoffs need a smart system for labeling}
\label{ssec:large_space}

Offline profiling over a set of calibration emails is an effective way to assign criteria for labeling (\eg when to choose the SLM cascade for labeling) However, naive profiling leads to prohibitive costs due to the large combinatorial tradeoff space, as there are several knobs which must be explored. Table \ref{tab:knobs} provides a summary of the knobs a profiling system needs to consider for making optimal tradeoff decisions.

Conducting an exhaustive sweep over all combinations of knobs significantly increases the cost of the profiler. In Figure \ref{fig:comb_cost}, we vary \emph{only} the confidence threshold for the SLM cascade in the range $80-100\%$, while keeping all other knobs fixed. The cascade uses three popular open-source SLMs \emph{microsoft/phi-4-mini-instruct}~\cite{microsoft_phi_4_mini_instruct}, \emph{meta-llama/Llama-3.1-8B-Instruct}~\cite{meta_llama_3_1_8b_instruct} and \emph{meta-llama/Llama-3.3-70B-Instruct}~\cite{meta_llama_3_3_70b_instruct}. 

We measure the quality (y-axis) using the F1-score and cost saved using the API cost reduction factor (x-axis) with respect to GPT4.1 We see that the cost-quality tradeoff space varies significantly varying a \emph{single knob} in a small neighborhood. Hence, it is \emph{essential} to find a method to reduce the space of profiling, to achieve a cost-efficient approach to reaching the best solution in the quality-cost tradeoff space.

\section{\name: A System for Cost-Efficient Enterprise Email Labeling}
\label{sec:design}

We present \name, an email importance labeling system for enterprise scale. \name consists of two main components - 

\begin{packeditemize}
    \item An \emph{efficient profiler} which runs offline profiling to determine the optimal values of knobs for the best quality-cost tradeoffs for email labeling (\S \ref{ssec:profiler}, \S \ref{ssec:adaptive_labeling} and \S \ref{ssec:cost}).
    \item An on-demand resource provisioning scheme to scale \name to minimize the cost increases during peak load. Services such as \anonemail are charged penalties for exceeding provisioned SLM capacities under load, making it essential to develop a solution to minimize this cost (\S \ref{ssec:provisioning})
\end{packeditemize}

\subsection{How and what do we profile?}
\label{ssec:profiler}

We showcase the working of \name's profiler in Figure \ref{fig:profiler}. As requirements of the email labeling workload, the profiler takes in the following inputs - 

\begin{packeditemize}
    \item \emph{Calibration set} - This is an input set of emails used by \name in the offline phase to decide factors such as which SLMs must be used in the cascade etc. as illustrated in Table \ref{tab:knobs}.
    \item \emph{Email Labels} - This is \emph{central} to \name's design and the tradeoff space. It is the list of labels with their requirements (\eg, nature of the label) which is used to determine the best labeling method to assign values at runtime.
    \item \emph{SLMs and embedding models} - This contains the list of SLMs and neural text embedding models which can be used in the final labeling pipeline.
    \item \emph{Range of confidence thresholds} - This provides lower and upper bounds of confidence thresholds, from which the SLM cascade must pick the best values for optimal quality-cost tradeoffs. On comparing with this value, \name can decide to keep the label from a given SLM or move up the cascade if is is below the threshold. The confidence threshold is computed by translating the \emph{log-probs} of the output tokens to a \emph{linear} confidence value for interpretability.
\end{packeditemize}

In the offline phase, \name's profiling is performed by measuring the quality-cost tradeoff of the choice of a given knob's value, with respect to the same label (and cost of generating it) generated by the LLM baseline. This profiling gives us \emph{Pareto-efficient} choices for the values of the knobs used by the system to label the emails (Figure \ref{fig:profiler}).

\subsection{Selecting knob values with \name's profiler}
\label{ssec:adaptive_labeling}

\begin{figure}[t]
    \centering
    \includegraphics[width=\linewidth]{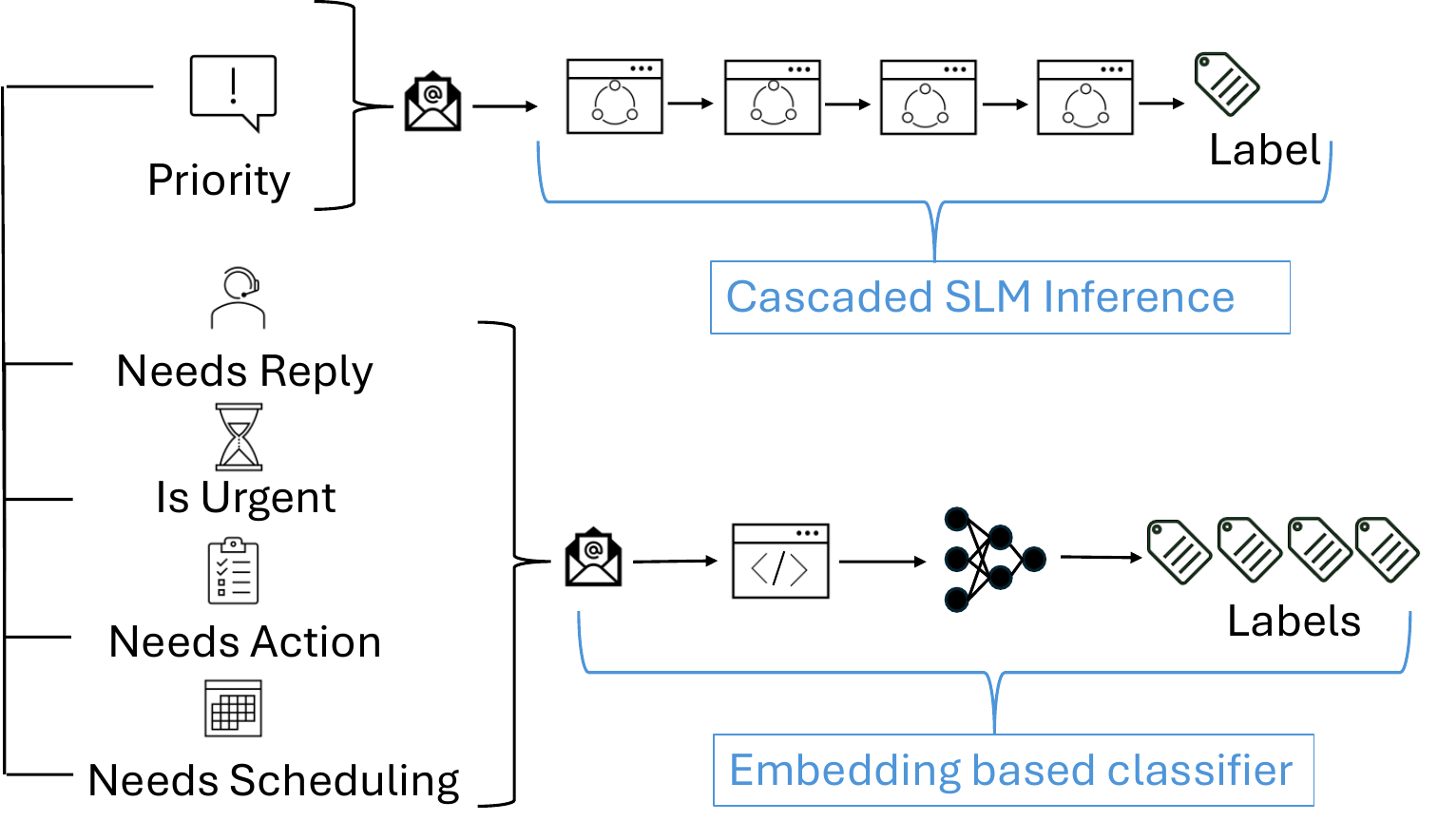}
    \vspace{-0.2cm}
    \tightcaption{\name decides labeling methods based on characteristics of the label and performance during profiling}
    \label{fig:hybrid}
\end{figure}

\name establishes that characteristics of the label can be used to choose the labeling method at runtime. Figure \ref{fig:hybrid} demonstrates the decision choice of using a labeling method. Based on extensive profiling experiments, \name \emph{decides} determines that an SLM cascade approach works well for the labeling, if the label is non-binary or multi-class in nature (\eg \emph{Priority}). If the label is a binary decision (\eg \emph{Needs Action}), a trained embedding based classifier suffices to decide the label value.

\begin{figure}[t]
    \centering
    \includegraphics[width=\linewidth]{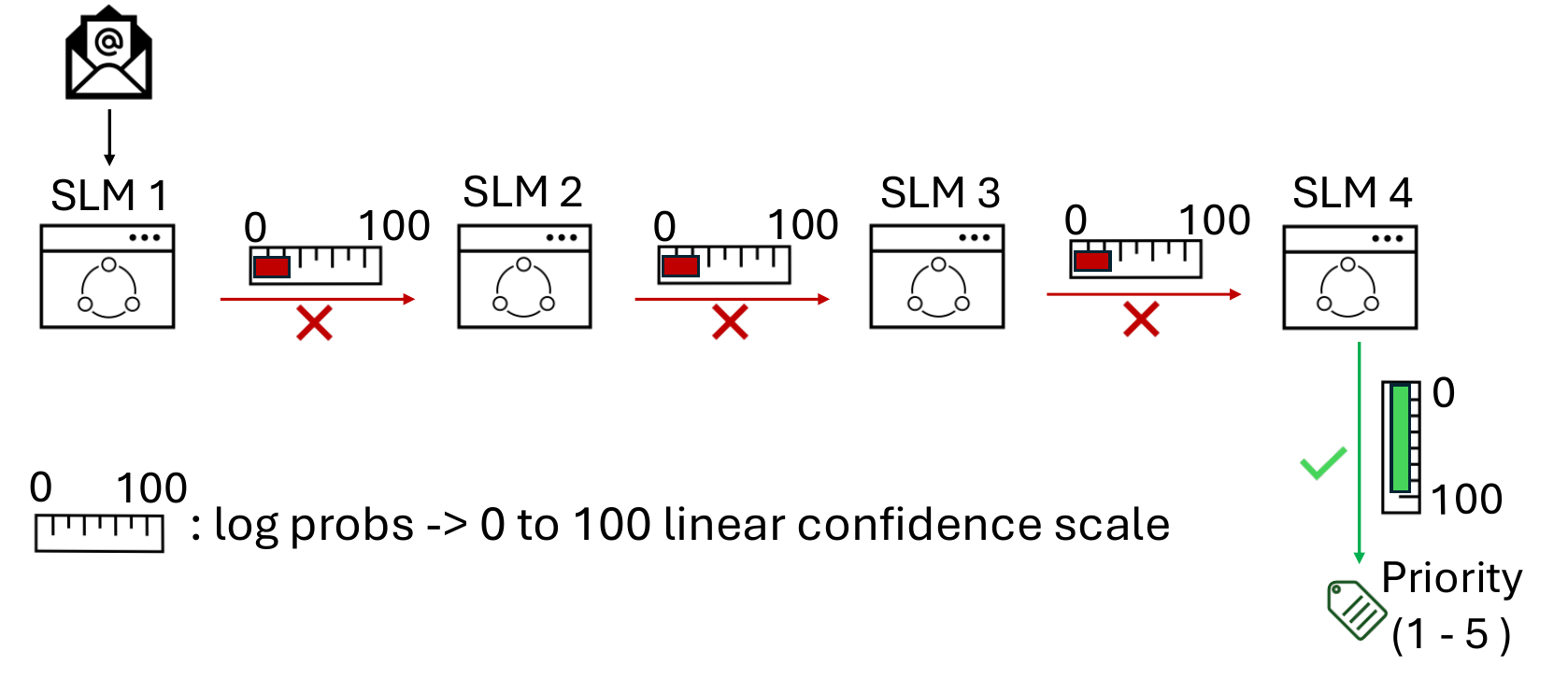}
    \vspace{-0.3cm}
    \tightcaption{Move up the cascade only if the previous models are insufficiently confident on the output value of the label}
    \label{fig:cascade}
\end{figure}

\mypara{Choosing knobs for the SLM Cascade}If \name chooses the SLM cascade for labeling, its profiling also provides the subset of SLMs and the confidence threshold values to be used in the cascade. The order of the SLMs determined by profiling is in increasing order of model size (parameter count) as this objective \emph{minimizes} the cost. We express details of the SLM cascade labeling method in Figure \ref{fig:cascade}, for assigning non-binary labels. Given an email, it is sent to the first model in the cascade. If the confidence of generating the label is above the chosen threshold, we keep the label, else the email is sent up the cascade and this process is repeated. 

The key insight is to use the cheaper SLMs for labeling the largest possible fractions of the emails and only use expensive SLMs when the cheaper models fail. Finally, if the end of the cascade is reached and no model has a confidence value higher than the thresholds , we keep the label generated by the \emph{most-expensive} model , as profiling shows larger SLMs have better quality in tail cases and the cost of running the model  has already been incurred.

\begin{figure}[t]
    \centering
    \includegraphics[width=\linewidth]{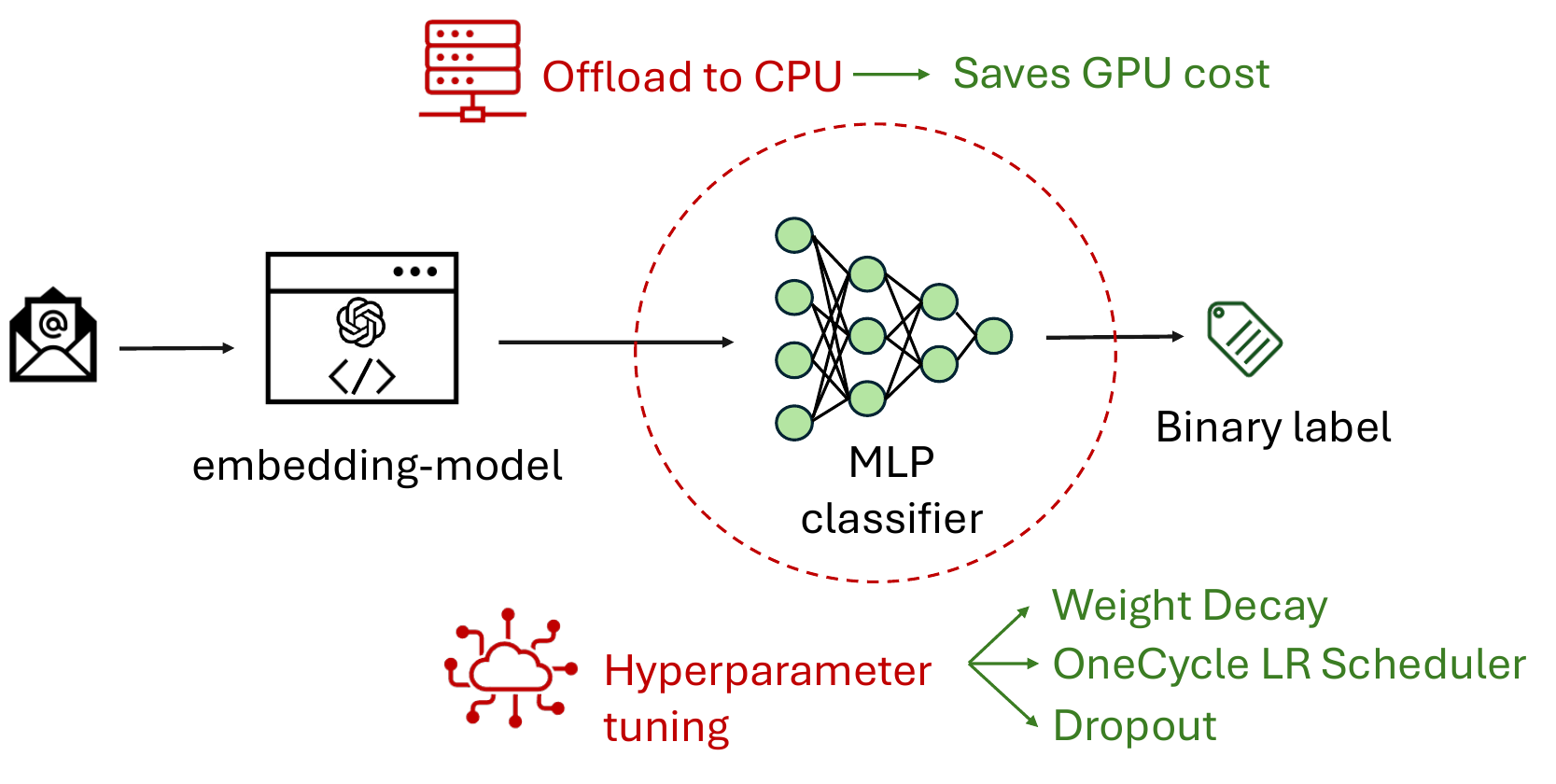}
    \vspace{-0.2cm}
    \tightcaption{Embedding classifier suffices for binary labeling}
    \label{fig:embedding}
\end{figure}

\mypara{Choosing knobs for the embedding classifier}We express the details of the embedding classifier labeling method in Figure \ref{fig:embedding} for binary labels. The classifier is 3-layer neural network and is trained offline using the frozen embeddings of the emails, from the same input calibration set used by the SLM cascade. The profiling phase provides the optimal choice of embedding model to be used in the pipeline. The generated embeddings are reused across all binary labels. Additionally, the classifier is completely offloaded to the CPU for training and inference , which saves on GPU compute costs. Further micro-optimizations in hyperparameter tuning such as weight decay~\cite{loshchilov2019decoupled}, onecycle learning rate scheduling~\cite{smith2017cyclical} and dropout~\cite{srivastava2014dropout} improves the classifier performance.

\begin{figure}[t]
    \centering
    \scalebox{0.78}{
        \includegraphics[width=\linewidth]{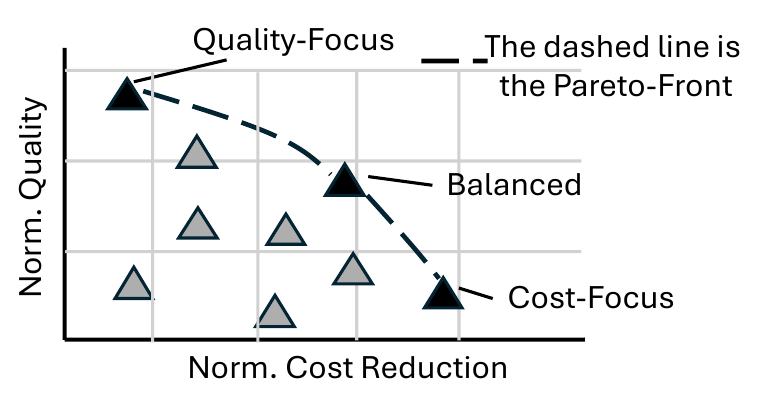}
    }
    \tightcaption{\name's default policy chooses the Balanced point}
    \label{fig:tradeoff}
\end{figure}

\mypara{Tradeoff Policy}\name's objective is to perform email labeling while balancing cost and quality. \name balances the tradeoff as shown in Figure \ref{fig:tradeoff} by choosing a point on the \emph{Pareto Front} which \emph{maximizes} sum of the \emph{normalized} generation quality (wrt. to baseline LLM) and the cost reduction (wrt. to cheapest model) . We measure quality using the F1-score between the labels generated by \name and the baseline LLM and cost reduction using the cascade's API cost reduction factor, with respect to the baseline LLM. By default, the \emph{Balanced} policy is chosen. However operators are provided an interface to specify weights for each dimension, to optimize for custom tradeoffs. We evaluate this in Section \ref{sec:eval}.

Additionally, it is important to note that \name's approach which incorporates \emph{SLM cascades} for labeling differs from a current line of work on \emph{model-routing}~\cite{ding2024hybridllm, hu2024routerbench, ong2024routellm, semanticrouter2024}, where inputs are routed to different models at runtime for inference, using a small classifier based on the \emph{input text}. \name does not make decisions based on the input email content at runtime as training such a classifier for email labels does not scale (as shown in \S \ref{ssec:embedding_only}) with email heterogeneity and volume. Instead \name decides the labeling method based on the offline profiling phase using label characteristics. This assigns the labeling method at runtime without relying on email content.

Finally, while labeling email workloads is chosen as a representative task for this work, \name's design can be used for labeling of any enterprise-scale messaging workloads, even beyond emails (\eg instant messaging), as our system does not make any specific assumptions constrained to emails.

\subsection{How do we manage the cost of profiling?}
\label{ssec:cost}

\mypara{Incrementally building the calibration email set}A natural challenge in \name's design is controlling the cost of the offline profiling, so as to actually realize cost savings in inference. The first dimension we consider is the size of the offline calibration set of emails for determining the values of the profiling knobs. If the size of the set is too small, the calibration of the knobs may not generalize while if the size is too large, the profiling may waste resources, without providing additional gain in the tradeoff space. \name tackles this by incrementally building the calibration set, with respect to a separate holdout \emph{validation set}. We start with a small calibration set, profile on it and test on the validation set. We continue increasing the size of the calibration set till \name finds points on the quality-cost \emph{Pareto Front} on the validation set. At this point, we can bound the size of the calibration set as further increases in size, does not improve the quality-cost tradeoff as the Pareto efficiency boundary has been reached.

\begin{figure}[t]
    \centering
    \includegraphics[width=\linewidth]{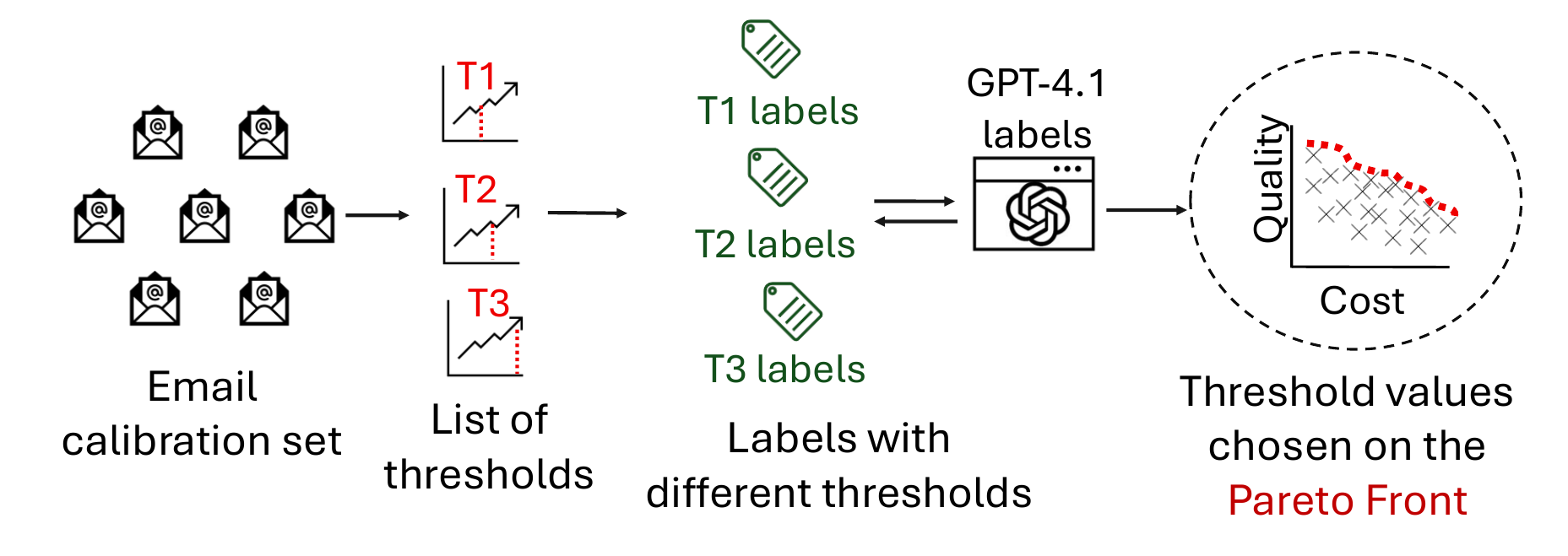}
    \vspace{-0.4cm}
    \tightcaption{Efficiently choosing confidence thresholds on the Pareto Front for use in the SLM cascade}
    \label{fig:pareto}
\end{figure}

\begin{figure}[t]
    \centering
    \includegraphics[width=\linewidth]{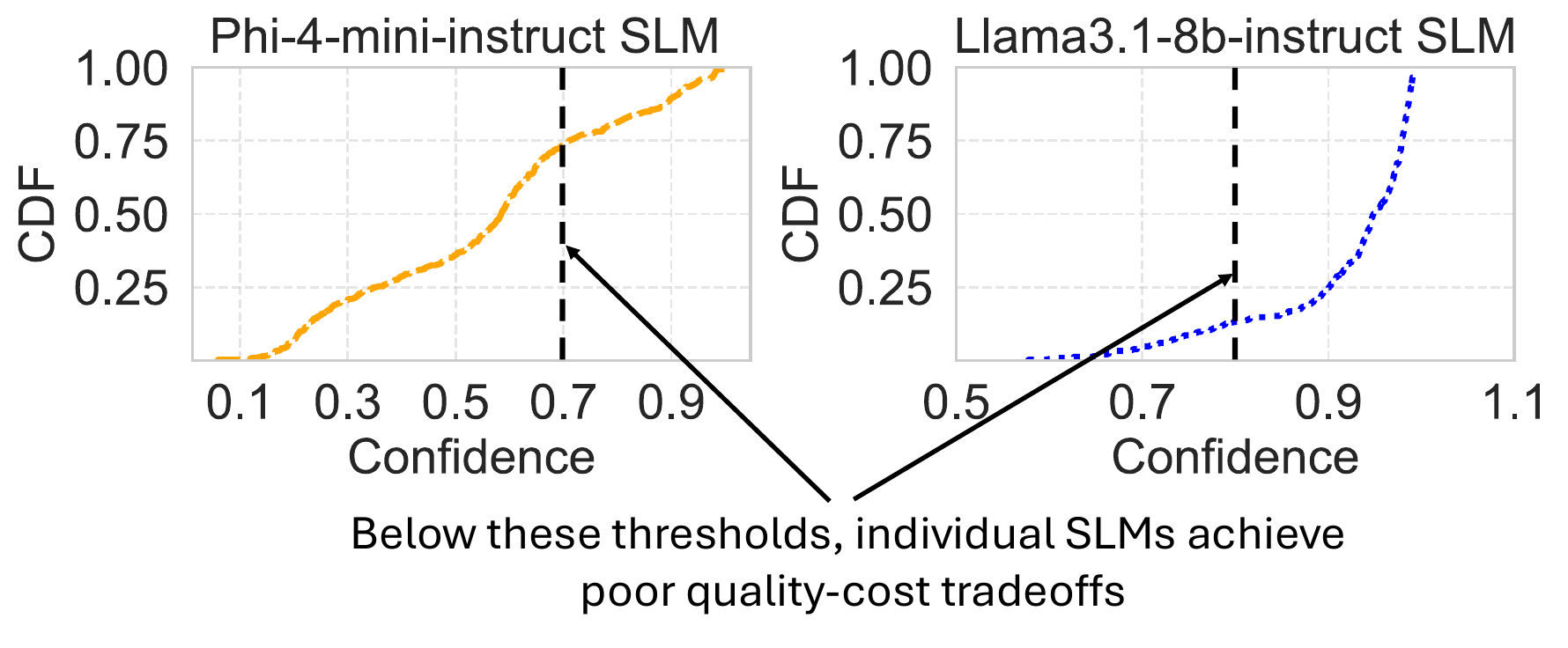}
    \vspace{-0.4cm}
    \tightcaption{Log-prob confidence distribution for two open-source SLMs on a calibration set from the Enron Dataset}
    \label{fig:confidence}
\end{figure}

\mypara{Using independence between profiling knobs} Another aspect \name utilizes to reduce the cost of the profiler is using the empirically observed independence among the different profiling knobs.The \emph{first} observation is that optimal SLM models to be used in the cascade can be independently chosen based on individual performance in the tradeoff space. The \emph{intuition} for this follows from every SLMs having its own quality-cost tradeoff, unaffected by others. Using SLMs with more optimal tradeoffs automatically forms a more optimal cascade. \emph{Second}, the objective is to minimize the cost of inference. Ordering of SLMs by increasing size minimizes this, as the goal is to use the cheapest models as frequently as possible and try cheaper models before expensive ones. As it is not possible to decide a priori which SLM in the cascade will be confident enough to label a given email, ordering by size will minimize the cost as \name does not use the input email content to make any model selection decision.

Finally, \name performs a offline sweep over a list of confidence threshold values as shown in Figure \ref{fig:pareto} across the SLMs in the cascade, to choose a subset of threshold values which lead to points on the \emph{Pareto Front} in the cost-quality tradeoff space. We observe that the input list of confidence thresholds does not need to be exhaustive. In Figure \ref{fig:confidence}, we observe that for two popular SLMs of \texttt{phi-4-mini-instruct} and \texttt{llama3.1-8b-instruct}, below confidence values of $70\%$ and $80\%$ respectively, the individual SLMs consistently achieve poor quality(disagreement with baseline LLM label).To reduce the cost of profiling over the list of thresholds, we can discard those for which the individual SLMs are already known to under-perform. This information is known during selection of the SLMs when building the cascade and reduces the sweep space for profiling different thresholds.

\begin{equation}
\begin{aligned}
\text{SWD} &= \frac{W_1(F_X, F_Y)}{\sigma_X}, \\
\text{where} \quad
W_1(F_X, F_Y) &= \int_0^1 \Big| F_X^{-1}(u) - F_Y^{-1}(u) \Big| \, du, \\
\text{and} \quad \sigma_X &= \mathrm{std}(X)
\end{aligned}
\label{eqn:sws}
\end{equation}

\mypara{Periodic and signal-driven re-profiling} As email content distributions may change over time, \name needs a component for re-profiling to accommodate these shifts, which makes it essential to have a \emph{low-cost} profiler. \name chooses performs re-profiling using \emph{golden labels} which are generated periodically using the baseline LLM. We use the Standardized Wasserstein-1 Distance (SWD)~\cite{he2024gradualDomainAdaptation} shown in Equation \ref{eqn:sws} to measure the distribution shifts in the log-prob confidence values of the SLM cascade. The profiling frequency is chosen based on a combination of two factors, (a) profiling once per day and (b) if there is a change $ > 1.0$ in the SWD~\cite{balaji2019normalizedwassersteindistancemixture} on the log-probs of the arriving emails, with respect to initial emails used by the profiler, whichever occurs first. We acknowledge that this heuristic can be further tuned for specific improvements if more domain-specific insights are available. Finally, \name's re-profiling signals can be further enhanced based on feedback on the labeling from the client (\eg several email clients offer this feature) or based on the user's profile (\eg designation, organization etc.) , and we leave these user-specific integrations to future work.

\subsection{On-demand provisioning to minimize email labeling costs under increasing load}
\label{ssec:provisioning}

\begin{algorithm}[t]
\caption{Greedy Allocation With Instance Penalty and Strict Running-Cost Ordering}
{\small
\KwIn{
    Total demand $D$, number of requests $r$, instance capacity $C$, \\
    instance base-costs $c_i$, penalty $p \ge 1$, \\
    running costs per request $z_1 < z_2 < \dots < z_m$, \\
    initial steady-state requests $k_i$, initial instances $n_i$, \\
    number of models $m$ in the cascade
}
\KwOut{Updated $n_i$, $k_i$, TotalCost}

% Per-request demand
$D_{\text{req}} \gets D / r$

% Process new incoming requests (demand growth)
\For{each new request}{
    \For{$i \gets 1$ \KwTo $m$}{
        \If{$k_i D_{\text{req}} < n_i C$}{
            $k_i \gets k_i + 1$        \tcp*{Model i has free instance capacity}
            \textbf{break}
        }
        \Else{
            % Decide: expand this model or use the next one
            $\Delta InstCost \gets c_i p^{\,n_i}$        \tcp*{Provisioning a new instance}
            $\Delta RunCost \gets z_{i+1} - z_i$         \tcp*{Cost of pushing to next model}

            \If{$\Delta InstCost < \Delta RunCost$}{
                $n_i \gets n_i + 1$      \tcp*{Cheaper to add an instance here}
                $k_i \gets k_i + 1$      \tcp*{Serve request on the same model}
                \textbf{break}
            }
            \Else{
                \textbf{continue}         \tcp*{Try the next more expensive model}
            }
        }
    }
}

% Final total cost
$\text{InstanceCost} \gets \sum_{i=1}^{m} c_i \frac{p^{n_i} - 1}{p - 1}$
$\text{RunCost} \gets \sum_{i=1}^{m} k_i z_i$
$\text{TotalCost} \gets \text{InstanceCost} + \text{RunCost}$

\Return{$n_i$, $k_i$, \text{TotalCost}}
}
\label{alg:provisioning}
\end{algorithm}

In addition to profiling and selecting methods for email labeling , \name also builds a resource-provisioning scheme to scale and manage the SLM cascade with increasing email load, by ensuring that serving such loads lead to minimal cost increases. \name focuses on provisioning for the SLM cascade as the embedding classifier does not require expensive hardware (runs on the CPU). Scaling the classifier with increasing load is simple as multiple instances with the same classifier weights can be loaded on demand on the different CPUs, which is $>100\times$ cheaper than the cheapest SLM.

Businesses have organizations like \anonemail, which can use SLM models for labeling, via \emph{model-as-a-service} hosted APIs and are billed per request based on real-time use. These SLMs are provisioned for a pre-determined capacity for the use by \anonemail organization, based on \emph{averaged} utilization trends. If operated at under capacity, they are only billed per-request. If they exceed the capacity, there is a penalty applied on provisioning further instances in addition to the per-request billing. This ensure services such as \anonemail do not get an unfair share of the \emph{model-as-a-service} APIs compared to other organizations in the overall business. Using this, we setup the cost model of the SLM cascade with two main costs for inference based on requirements of \anonemail when deploying such labeling solutions.

\begin{packeditemize}
    \item \emph{Per-request usage cost} - This is the inference cost for running a request for a given model, based on the input and output token lengths. As both of these lengths are known, this cost is easily measurable. Requests can be sent to the instance of the model till it reaches capacity and are only billed when the instance is used to label the request.
    \item \emph{Model instance provisioning cost} - This is the cost to create an instance for a given model in the SLM. Initially, instances for each model are created based on the individual model utilization rate measured during the profiling phase. Provisioning instances when the system is under capacity only incurs the base cost. If capacity for the current instance is reached, every additional instance for a model has a multiplicative penalty factor $p$, due to resource limitation and ensure fair sharing with other services.
\end{packeditemize}

We assume the inference pipeline can provision additional instances readily, as they are available in a pre-provisioned pool and can be accessed via a \emph{model-as-a-service API}. We aim to minimize the cost increase of the SLM cascade when capacity is reached for the current allocation, as shown in Algorithm 1. We start with an initial provisioning allocation for each model with fixed base costs $c_{1},\dots,c_{m}$. The inference cost of the requests $z_1 < z_2 < \dots < z_m$ increase up the cascade as these costs are proportional to the model size, and the cascade is ordered by increasing size.

Algorithm 1 minimizes cost by minimizing the \emph{marginal cost increase} at every cost increase decision step. In lines $4-6$, it first assigns requests to the current model instance in the cascade, as long as it has capacity left. If the email arrival rate increases and the capacity for an instance is reached, in lines $8-15$, \name measures the marginal cost increase of both provisioning a new instance with penalty $p$ or moving up the cascade to a model with higher \emph{per-request cost}. Based on whichever decision adds lower \emph{marginal cost}, it decides to add a new instance or move up the cascade, and it continues this process for the subsequent requests.

Algorithm 1 is \emph{optimal} for minimizing cost increases for inference. This is due to (a) the per-request costs are monotonically increasing up the cascade and (b) at every step when a cost increase decision is made, we minimize the \emph{marginal cost} increase. As costs are additive, minimizing the increase at each step, ensures the total cost is minimized. Furthermore, as moving up the cascade strictly increases the cost at every step, it is enough to consider moving up just one step in the cascade for the marginal cost calculation, as any additional move up steps , will further increase the cost.

\mypara{Organization policy can influence email labeling decisions}In addition to minimizing the cost increases of labeling during high email load, \name allows use of supporting information from email service to enforce selective policies for labeling. Organizations using large email volumes for business communication often have hierarchies, groups and employee levels. Such information is directly available (\eg designation) in their contact cards (\eg C-suite, senior manager) and is \emph{logged by the email service used in enterprises}.

\name uses information about a user level or group to enforce cost reduction policies during peak load. It allows two policies, a \emph{quality-based policy} and a \emph{delay-based policy}. For the quality-based policy, \name allows downgrading to a cheaper SLM in the cascade for non-quality sensitive groups (by overriding the SLM chosen by \name's profiler) during peak load. This reduces the increased cost incurred from the penalty factor by exceeding the instance capacity, albeit at some degradation in labeling quality. The choice of SLM for downgrade depends on both the difference in cost-quality values identified in the profiling phase and business policy.

For the delay-based policy, \name allows delaying the labeling of a given email, when the instance required to label it reaches peak capacity. The allowed delay is proportional to the duration of the increased load (while at capacity) and the enforcement is chosen based on the user's level or group. The delay reduces the labeling cost by \emph{staggering the requests}, which prevents the instance hitting capacity and in turn does not incur the penalty factor for newer instance provisioning.

Based on conversation about the business requirements of \anonemail, such policies may be rolled out based on the designations, levels and budgets of different groups. Given that additional cost increases are minimized, \emph{tolerable} drop (which is identified in the profiling phase by \name) in quality and increased delay are acceptable for certain user groups.

\tightsection{Enhancements to \name}
\vspace{0.1cm}
\label{sec:enhancements}

\name is developed to be a robust and self-sufficient system for importance labeling in enterprise email workloads. However, we want to allow flexibility in adapting the system to new requirements and capabilities.

\mypara{Enforcing external constraints for labeling}By default, \name aims to provide the optimal values for the profiling knobs in order to achieve the best quality-cost tradeoffs, universally over the entire possible input space. However, operators such as \anonemail might often want to enforce constraints in the system in order to comply with additional, external policies (\eg data sensitivity, only use on-prem hosted models etc.) \name easily enables such constraints. 

Before each offline profiling phase, \name provides an text-based interface to constraint the input space across all of the profiling knobs. One such use-case we measure in the current scope is only allowing a operator-defined set of SLMs to be used in the cascade (details in Section \ref{sec:eval}). Such a feature is useful for services such as \anonemail , which arise out off business policies or privacy constraints. \name considers all such constraints in the input space and run the offline profiler aware of them, in order to provide the optimal knob values within the reduced space of constraints.

\mypara{Beyond using off-the-shelf models}The standard \name profiler chooses from a pool of open-source SLMs to build the SLM cascade. It can for useful an enterprise to have its own fine-tuned model for use in the cascade, due to organization-specific email and data which could be used to achieve better cost-quality tradeoffs. \name provides additional interfaces to fine-tune SLMs and classifier models further, in order to incorporate custom-knowledge in the labeling pipeline. This reduces the need to choose between adaptivity and customization for labeling, by allowing to both techniques in the system. Fine-tuned models can be incorporated in the solution space, similar to enforcing constraints as above.

\mypara{Summarize emails beyond numeric labeling} While email importance labeling boosts productivity and efficiency, it helps further by adding features beyond numeric labeling. \name enables a feature of \emph{email summarization}, which may enable additional interpretability of an email's importance. As before, \name's profiler provides the subset of SLMs to use for the summarization task, the confidence thresholds and chooses the appropriate SLM at runtime to generate the summary. The profiler module can be used \emph{as-is} for summarization, with minor modifications in the decision metrics, with the goal remaining to match the output quality of the GPT4.1 LLM at lowest possible cost.

Minor modifications include using frequency-weighted confidence scores at the token level for choosing the appropriate SLM in the cascade, as confidence thresholds score become noisier over longer sequences. It also switches to using an \emph{LLM as a judge} to evaluate the quality of the summarized outputs as F1-score also becomes noisier over longer decoded sequences. The core system of \name remains the same for enabling this additional feature and we evaluate the performance of email summarization in Section \ref{sec:eval}.
\tightsection{Implementation}
\vspace{0.1cm}
\label{sec:implementation}

We implement \name in Python. The core implementation has two components - the \emph{profiler module} and the \emph{resource provisioning module}. \name's goal is to develop an email labeling scheme for enterprise-scale, hence several implementation detail choices and cost abstractions are made keeping in mind the ease-of-deployment for existing email services like \anonemail. The primary bottleneck is the labeling cost and SLM API availability. The email workload are less sensitive to traditional SLM metrics such as latency and time-to-first-token (TTFT). Metrics like GPU utilization and compute are contained within the \emph{cost} metric of the SLM or classifier.

The SLM cascade can be used in two modes. The first uses either vLLM~\cite{kwon2023efficient} or Huggingface Transformers~\cite{wolf-etal-2020-transformers} to run open-source SLM instances on on-prem GPUs and the second uses SLMs hosted via API services on Azure's AI Foundry~\cite{microsoft2025azureai}. For the resource provisioning scheme, \name currently uses SLM APIs in a \emph{model-as-a-service} mode hosted on Azure's AI Foundry, as it provides easy integration with current \anonemail code. This ensures availability of SLM endpoints to match real-time demand, without inhibiting \anonemail by instance spin-up times and GPU lease times.

The embedding classifier is implemented in PyTorch~\cite{paszke2019pytorch} and PyTorch Lightning~\cite{falcon2025pytorchlightning}, with support for running the model in the industry standard ONNX-weight format~\cite{onnx2025}, allowing easy deployment of ML models and pipelines. We use the Adam optimizer~\cite{kingma2015adam} with its suggested parameters of $\beta_1 = 0.9$, $\beta_2 = 0.98$, and $\epsilon = 10^{-9}$, one-cycle learning rate with an maximum value of $5e^{-4}$ with cosine annealing and weight decay of $1e^{-5}$. \name can generate email embeddings locally on the CPU/GPU servers via an integration with SentenceTransformers~\cite{reimers-gurevych-2019-sentence} or via hosted services with OpenAI APIs~\cite{openai2020api} and Azure's AI Foundry. Finally, \name has text-based interfaces of \texttt{enforce_constraints} and \texttt{available_model_pool} to supply operator-defined requirements to be used for profiling and labeling inference.

\section{Evaluation}
\label{sec:eval}

\begin{figure*}[t]
    \centering
    \includegraphics[width=0.99\textwidth]{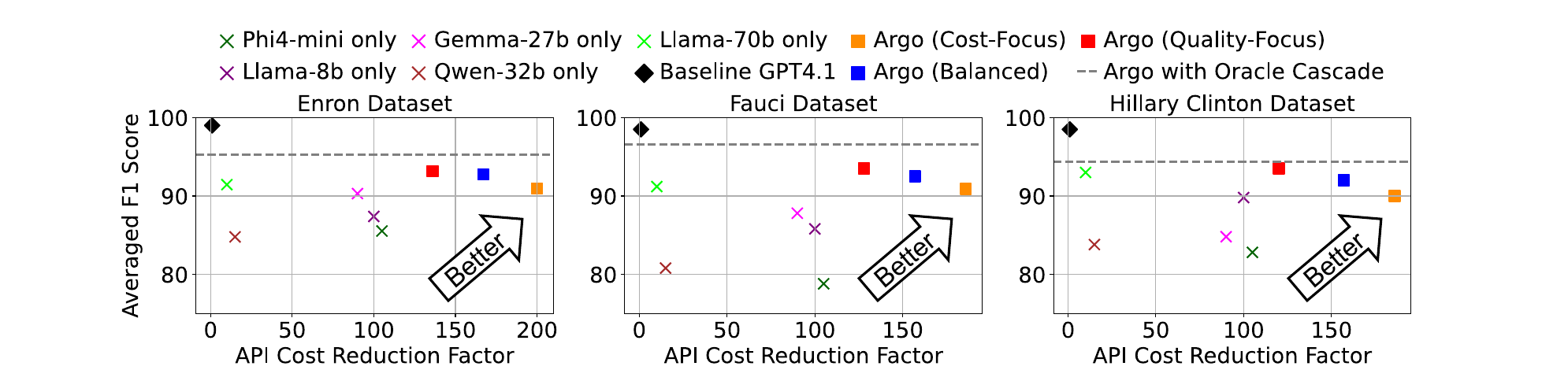}
    \vspace{-0.3cm}
    \tightcaption{\name finds multiple Pareto Efficient cost-quality tradeoff points for the SLM cascade and chooses the \emph{balance} point (blue square) which achieves 148-167$\times$ cost reduction with comparable quality}
    \label{fig:eval_e2e}
\end{figure*}

The key takeaways from the evaluation are

\begin{packeditemize}
    \item \name achieves 148-167$\times$ lower inference costs compared to the GPT4.1 baseline LLM at comparable quality.
    \item \name's resource provision achieves 2.2-3.8$\times$ lower cost increases compared to the baselines.
    \item \name's profiler achieves 20-640000$\times$ times lower profiling cost without any drop in quality compared to all baselines.
    \item \name is extensible, with pluggable components to incorporate needs and constraints of \anonemail like operators.
\end{packeditemize}

\subsection{Setup}
\label{ssec:setup}

\begin{table}[t]
    \centering
    \scalebox{0.78}{
    \begin{tabular}{l  l  l}
        \toprule
          Model Name & Type & API Cost Reduction  \\
         \cmidrule{1-3}
         microsoft/phi-4-mini-instruct~\cite{microsoft_phi_4_mini_instruct} & SLM & 105$\times$ cheaper \\
         meta-llama/Llama-3.1-8B-Instruct~\cite{meta_llama_3_1_8b_instruct} & SLM & 100$\times$ cheaper \\
         google/gemma-3-27b-it~\cite{google_gemma_3_27b_it} & SLM & 90$\times$ cheaper \\
         Qwen/Qwen2.5-32B-Instruct~\cite{qwen_2_5_32b_instruct} & SLM & 20$\times$ cheaper \\
         meta-llama/Llama-3.3-70B-Instruct~\cite{meta_llama_3_3_70b_instruct} & SLM & 10$\times$ cheaper \\
         text-embedding-3-large~\cite{text_embedding_3_large} & Embedding & 100$\times$ cheaper \\
         \bottomrule
    \end{tabular}
    }
    \caption{Summary of SLMs and embedding models chosen by \name's profiler to use for labeling}
    \vspace{-0.8cm}
    \label{tab:table_models}
\end{table}

\mypara{Models, Hardware and Software}\name's profiler used 5 open-source models for the SLM cascade and an embedding model for the classifier (summarized in Table \ref{tab:table_models}). All models are instruction-tuned and can take long contexts as input (upto 128K tokens). The API cost reduction factor is measured as the ratio of the standard blended token cost (3:1 input-output tokens)~\cite{openaiPricing2025} of running the models on Azure AI Foundry and the baseline GPT4.1 on the same input emails.

We use an NVIDIA A100 GPU server hosted on Azure, with upto 4 GPUs of 80GB HBM (each SLM needs different numbers of GPUs) to run the SLM cascade using either vLLM or Huggingface. The servers are equipped with 216GB of memory and an AMD EPYC 7v13 64-Core Processor, with 24 cores per socket. The embedding classifier is trained and run on a CPU server with the same specifications, achieving an inference time of less than $1ms$ per email. 

The \emph{model-as-a-service} mode is run on Azure AI Foundry using the OpenAI interface, as is the GPT4.1 LLM baseline.

\mypara{Datasets}We use 3 open-source email datasets for evaluating \name's cost-quality tradeoffs and performance.

\begin{packeditemize}

    \item Enron email dataset~\cite{enronemail} - The Enron dataset is public corporate email dataset consisting of 40000 emails from 137 inboxes of employees (senior management) of Enron. It has been widely used for research in email classification.

    \item Fauci email dataset~\cite{fauciemail2021} - A dataset of emails released in 2021 sent by Anthony Fauci documenting U.S. government correspondence related to the COVID-19 pandemic response. The dataset comprises of 2,761 emails and includes metadata like recipients, CC lists, subjects and timestamps.

    \item Hillary Clinton dataset~\cite{clintonemailarchive2016} - A public archive containing around 7000 emails and attachments sent to and from Hillary Clinton’s private email server during her tenure as U.S. Secretary of State (2010–2014). The archive has been used in linguistic and text-corpus research.
\end{packeditemize}

For the evaluation sets, we construct random samples of emails from each of the datasets of size 1000, 500 and 700 from the Enron, Fauci and Hillary Clinton datasets. The calibration sets used for profiling is from the \emph{remaining datasets} with no \emph{cross-contamination} with the evaluation set. For evaluation, classifier training and profiler calibration, the emails are labeled with the baseline GPT4.1 LLM. We measure inference quality-cost tradeoffs and profiler cost savings with \name.

\mypara{Inference Baselines}
\begin{packeditemize}
    \item GPT4.1 LLM~\cite{openai2024gpt4technicalreport}  - We compare the quality-cost tradeoff of labeling by using the F1-score of the generated labels by \name compared to the GPT4.1LLM baseline. As this baseline is prohibitively expensive to run by the \anonemail we have corresponded with, \name's goal is to maximize the cost reduction as least possible quality drop.
    chooses the label from the most expensive SLM in the cascade.
    \item Individual SLMs - We compare \name with the individual performance of every SLM summarized in Table \ref{tab:table_models}.
    \item Oracle SLM Cascade - We use \name with an \emph{oracle} SLM cascade as a reference line. This oracle has perfect knowledge and always chooses the SLM which provides the same label as the GPT4.1 baseline. If no SLM in the cascade agrees with the GPT4.1 label, the oracle chooses the label from the \emph{most-expensive} label in the cascade. The oracle serves as an upper-bound on the achievable quality. 
\end{packeditemize}

\mypara{Profiler Baselines}
    \begin{packeditemize}
    \item Exhaustive - This profiler sweeps over all profiler configuration knobs from the entire email calibration set.
    \item 1\% Sample - This profiler sweeps over all profiler configuration knobs from a random 1\% email calibration set.
    \item 10\% Sample - This profiler sweeps over all profiler configuration knobs from a random 10\% email calibration set.
    \item Reduced Cascade - This profiler only profiles with chosen 3 SLMs for the cascade with best individual quality-cost tradeoffs on the full calibration set.
    \item Reduced Thresholds - This profiler only profiles above median thresholds (50-100\%) on the full calibration set.
\end{packeditemize}

\subsection{End-to-End Results}

\begin{figure}[t]
    \centering
    \includegraphics[width=\linewidth]{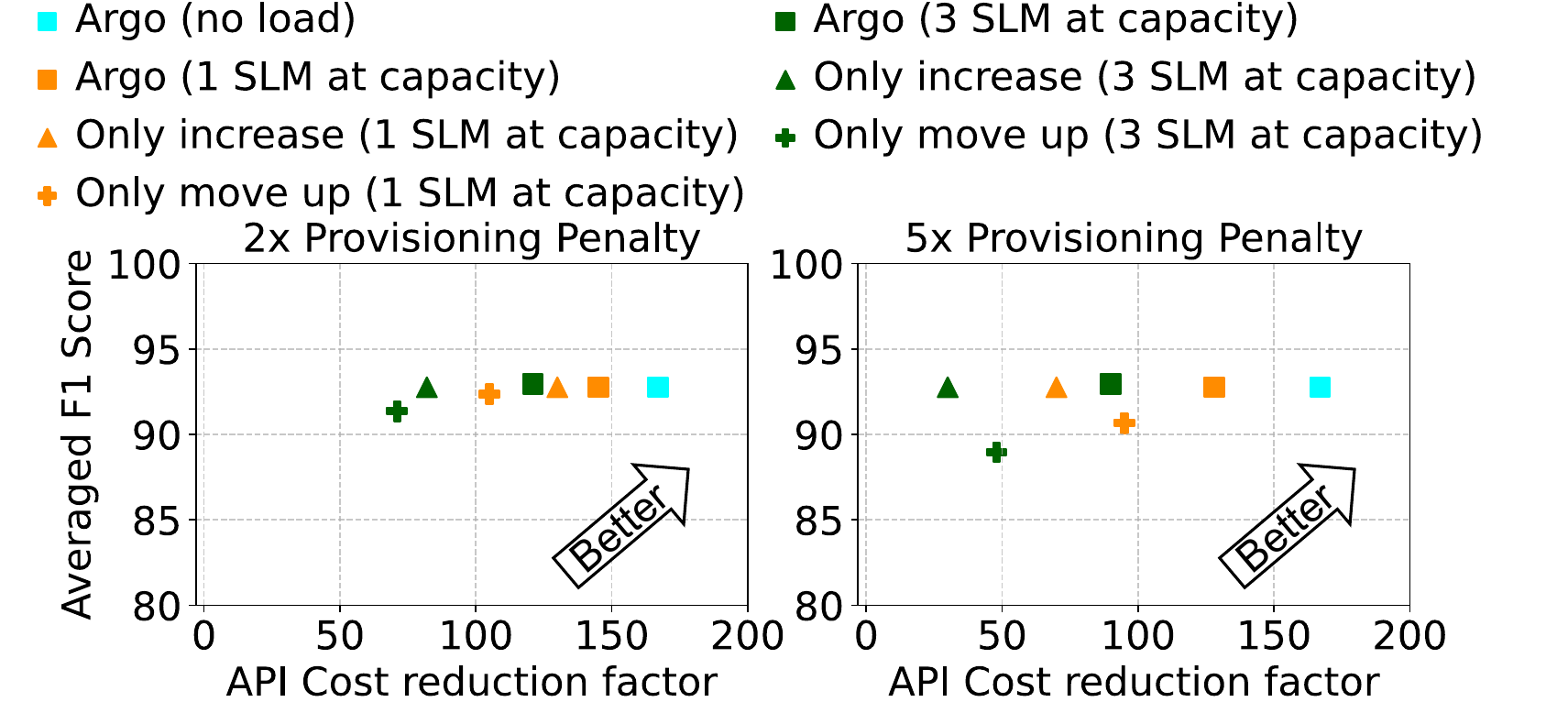}
    \vspace{-0.2cm}
    \tightcaption{\name resource provisioning is 2.2-3.8$\times$ lower cost than baselines across multiple penalty factors and capacity bottlenecks (results on Enron Dataset)}
    \label{fig:load}
\end{figure}

\begin{figure}[t]
    \centering
    \includegraphics[width=0.99\linewidth]{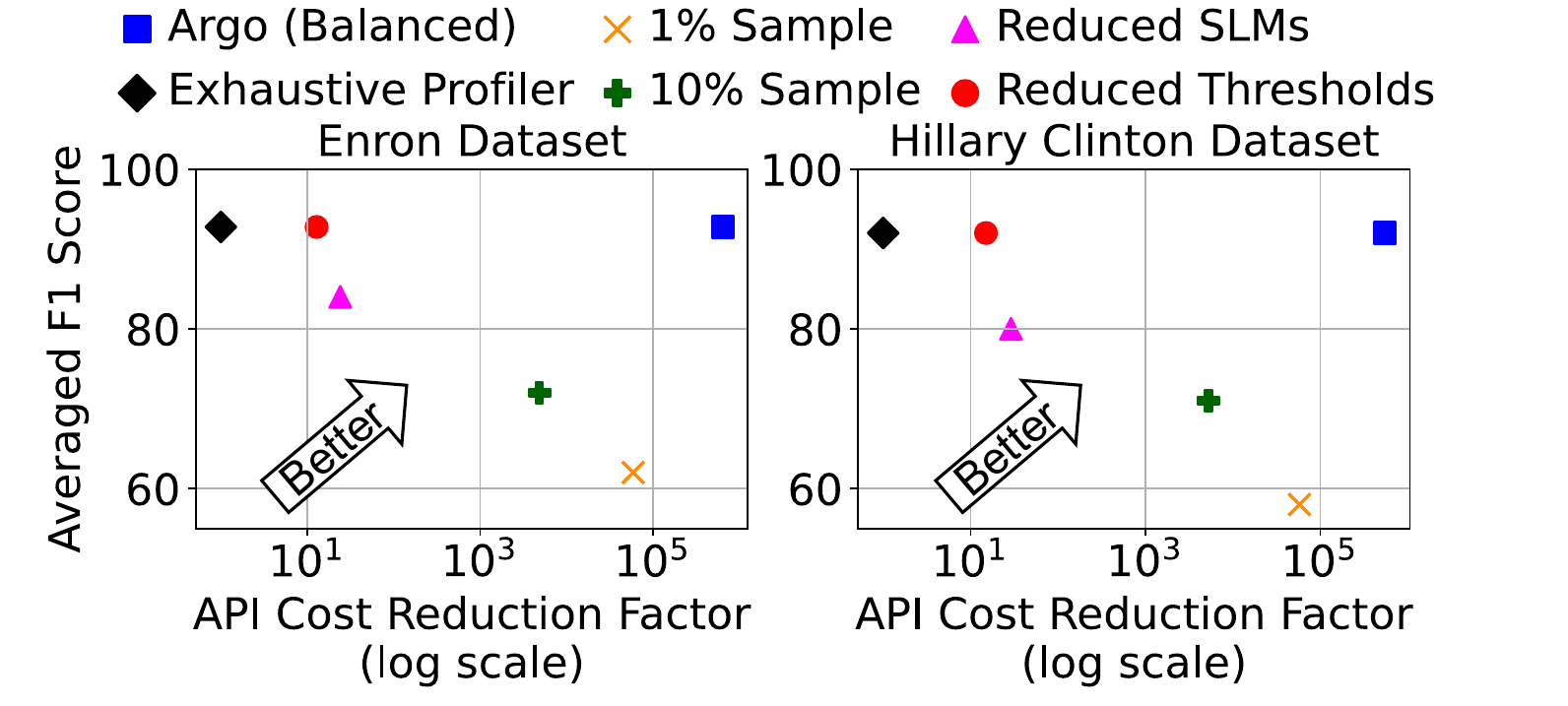}
    \tightcaption{\name achieves 20-640000$\times$ lower profiling costs compared to all baselines maintaining same or higher quality}
    \label{fig:profilercost}
\end{figure}

\begin{figure}[t]
    \centering
    \includegraphics[width=0.95\linewidth]{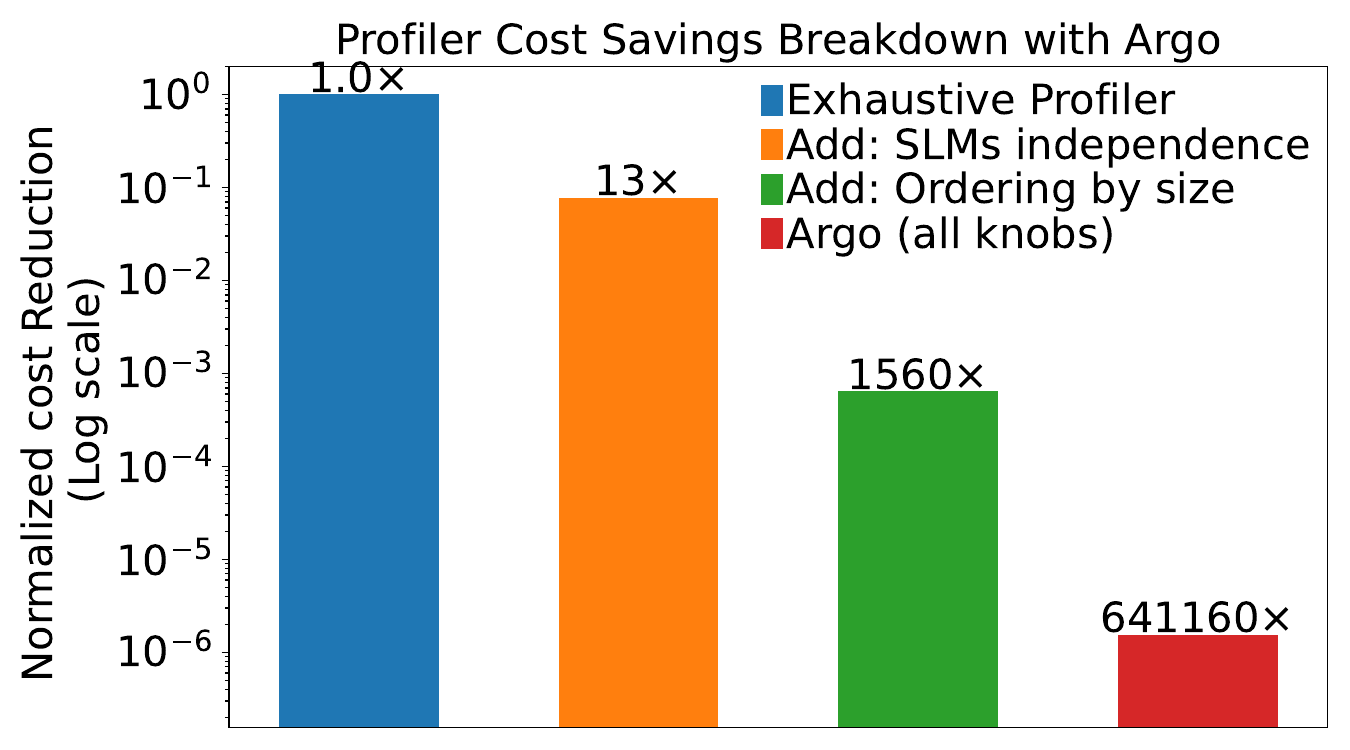}
    \tightcaption{\name reduces profiling costs by shrinking the search space using workload-space aware characteristics}
    \label{fig:cost_breakdown}
\end{figure}

\mypara{Cost-reduction at comparable quality} We show the overall improvements \name attains in improved quality-cost tradeoffs for email labeling in Figure \ref{fig:eval_e2e} on three open-source email datasets. As the system contains multiple labels, we report the average F1-score across all the labels. By default, \name chooses \emph{Balanced} which achieves 148-167$\times$ reduction in API cost as compared to the GPT4.1 baseline at comparable quality. \name achieves superior performance compared to every individual SLM in the cascade in both dimensions of better quality with 5-15\% higher quality and lower API cost by 1.67-17$\times$.

We also report two other choices \name provides , namely \emph{Cost Focus} and \emph{Quality-Focus} which also lie on the Pareto Front and achieve 186-200$\times$ and 120-136$\times$ reduction in API costs respectively and can be chosen if particular requirements need to be met for a given email workload. Finally, for our evaluation, quality tolerance is also validated using values acceptable by the business requirements of \anonemail.

\mypara{Cost-Efficient Scaling under peak load}We simulate load by increasing the arrival rate using the set of evaluation emails. We maintain a \emph{request-queue} for each SLM instance in the cascade and \emph{at-capacity} is reached when the queue is full. The initial allocation of model instances is based on the usage fraction of each SLM measured in the profiling phase. 

Figure \ref{fig:load} evaluates \name's provisioning under increasing load on the Enron dataset. We consider bottleneck capacities for 1 SLM (in orange) and 3 SLMs (in green) for two penalty factors of $2\times$ and $5\times$ as described in Section \ref{ssec:provisioning}. Across multiple penalty factors and capacity bottlenecks, \name achieves 2.2-3.8$\times$ lower cost increase as opposed to both baseline methods, (a) always increasing instances on demand and (b) always moving up the cascade for residual capacity.

\mypara{\name enables cheaper profiling without quality drop} In Figure \ref{fig:profilercost}, we compare \name's profiler cost against profiler baselines and measure the inference quality against GPT4.1 labels. \name achieves the same quality as the exhaustive profiler but is 640000$\times$ cheaper. The $1\%$ and $10\%$ sample baselines, while closer to \name in cost (31000$\times$ and 4850$\times$ cheaper than exhaustive respectively), achieve 33\% and 21\% lower quality respectively. The Reduced Cascade has a lower quality drop (12\%) but is only 24-29$\times$ cheaper than exhaustive profiling while the Reduced Threshold achieves the same quality but is only 12-15$\times$ cheaper than exhaustive profiling.

\mypara{Understanding \name's inference gains}\name reduces inference costs by selecting the cheapest labeling method which achieves comparable quality to the baseline LLM quality. Using the offline profiled results, \name chooses \emph{Pareto-Efficient} knob values in the inference pipeline and achieves 148-167$\times$ cost reduction at comparable quality. 

\subsection{Breakdown and Sensitivity Results}
\label{ssec:breakdown}

\mypara{Understanding the profiler cost savings}
In Figure \ref{fig:cost_breakdown}, we show the principled cost-reduction achieved by \name by \emph{smartly} choosing for each profiler knob. We compare  with the baseline cost of the exhaustive profiler on the calibration set, which does not use any insights from the workload characteristics. We \emph{incrementally} show the benefit from each profiling knob, using independence of knobs to reduce the search space and pruning values which have poor quality-cost tradeoffs. \name achieves upto 640000$\times$ reduction in the profiler cost as compared to the exhaustive profiler.

\begin{figure}[t]
    \centering
    \includegraphics[width=0.99\linewidth]{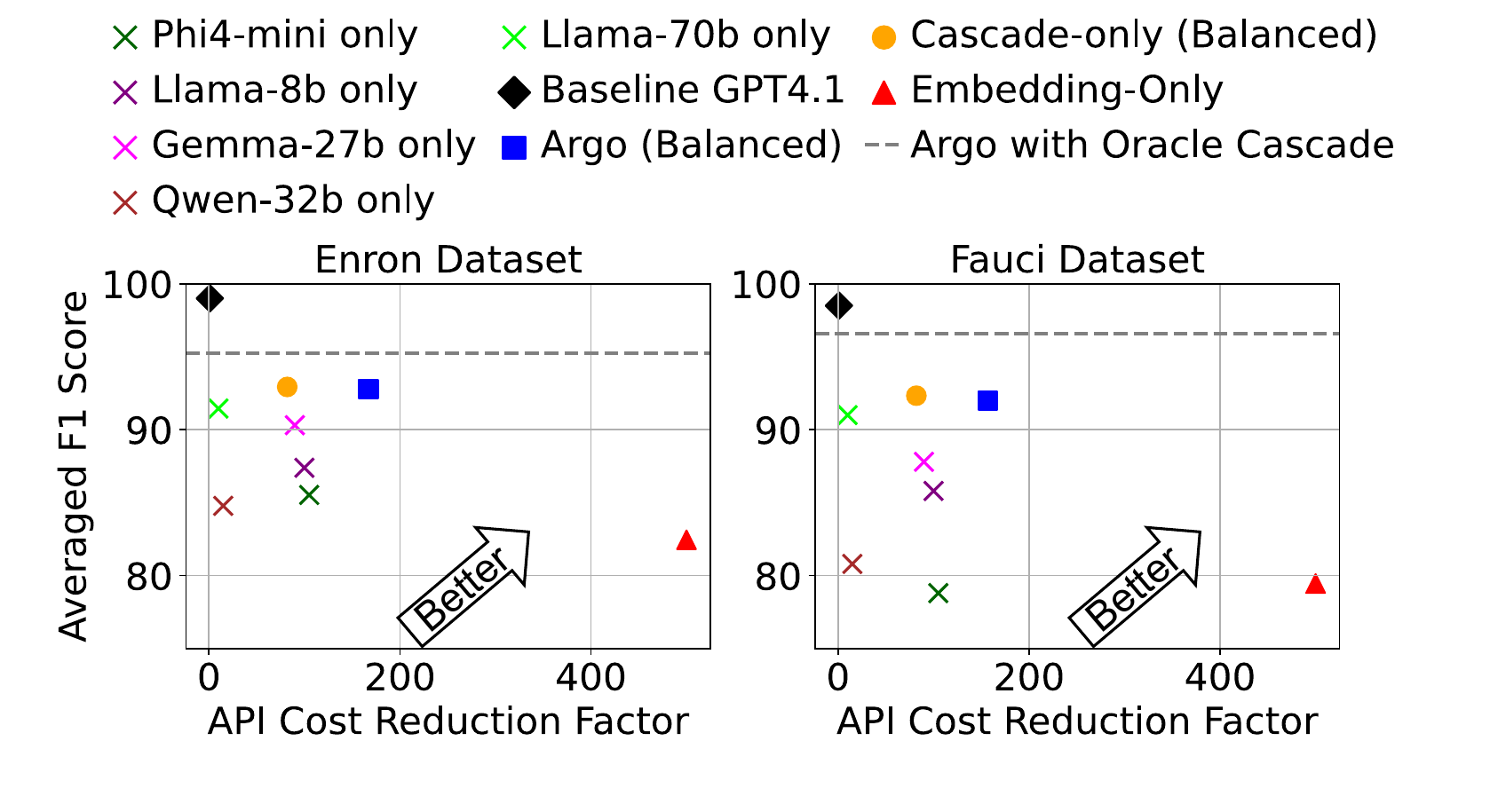}
    \vspace{-0.4cm}
    \tightcaption{Only using the SLM cascade or embedding classifier achieves 2$\times$ higher cost or 13-15\% lower quality}
    \label{fig:ab}
\end{figure}

\begin{figure}[t]
    \centering
    \includegraphics[width=0.99\linewidth]{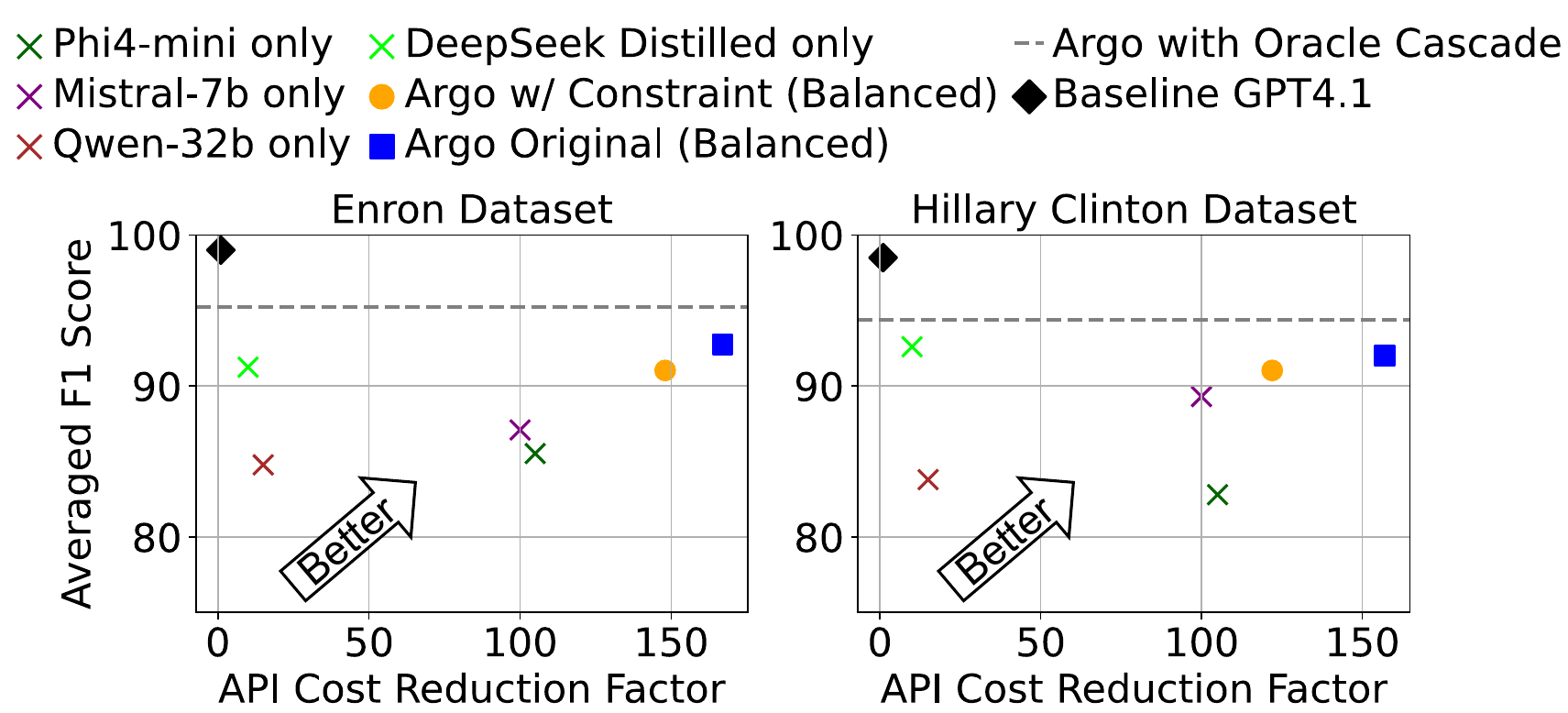}
    \tightcaption{\name enables specifying operator constraints (\eg SLM subsets to be used) and provides a tradeoff solution which enforces the constraints}
    \label{fig:constrained}
\end{figure}

\mypara{Why not only SLM or only embedding classifiers?}
In Figure \ref{fig:ab}, we show an ablation study with \name, where (a) we always use the SLM cascade for generating each label and (b) we always use the trained embedding classifier for it. If we only use the SLM cascade, the quality is comparable but we lose 2$\times$ on cost savings. If we only use embedding classifier, the quality drops by 13-15\%, which is unacceptable by services as \anonemail to deploy in production.

\begin{figure}[t]
    \centering
    \includegraphics[width=0.99\linewidth]{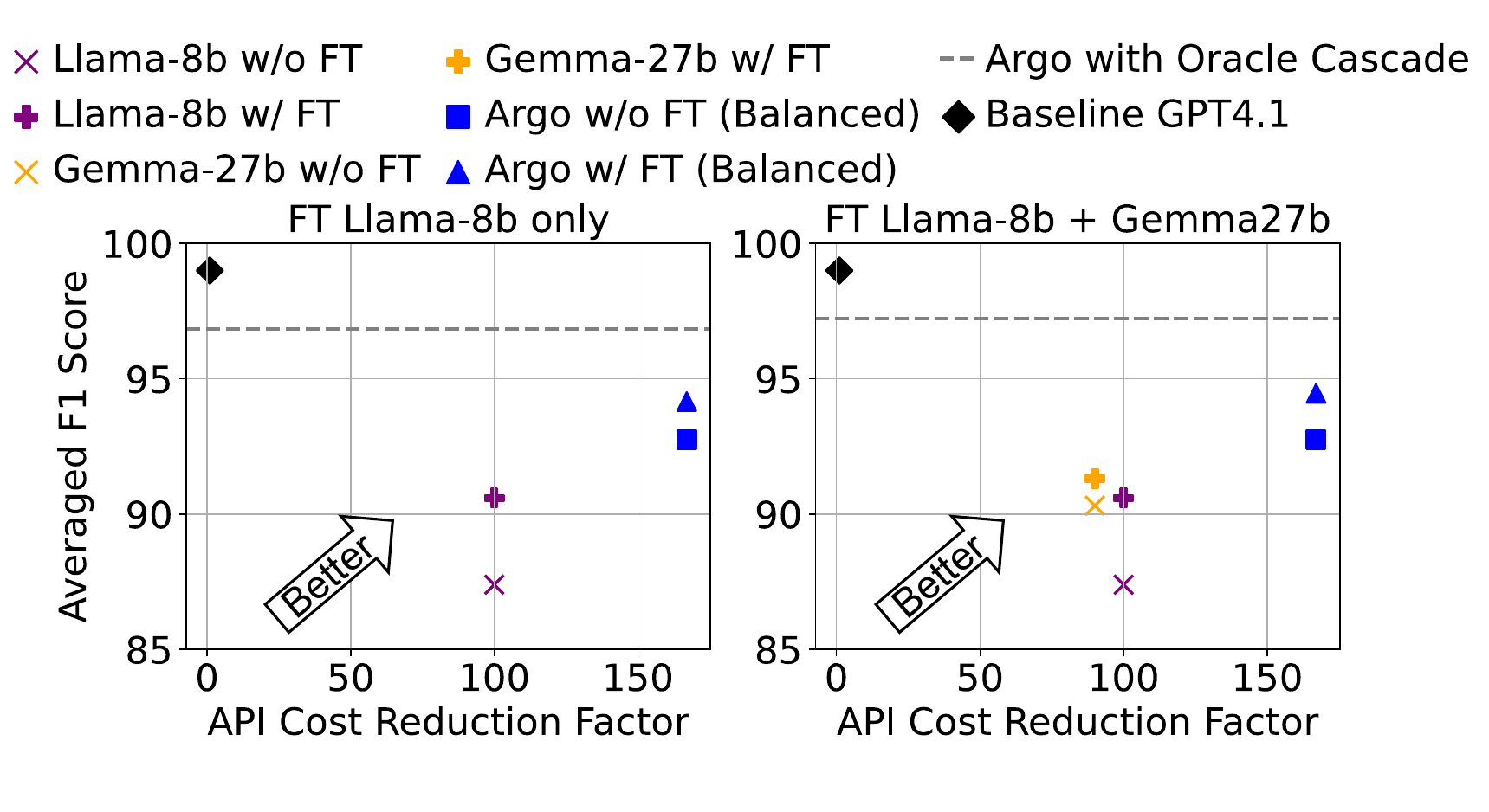}
    \vspace{-0.4cm}
    \tightcaption{Fine-tuning individual SLMs themselves does not provide better quality-cost tradeoffs than \name but can be added to \name for improvements (on Enron Dataset)}
    \label{fig:ft}
\end{figure}

\begin{figure}[t]
    \centering
    \includegraphics[width=0.99\linewidth]{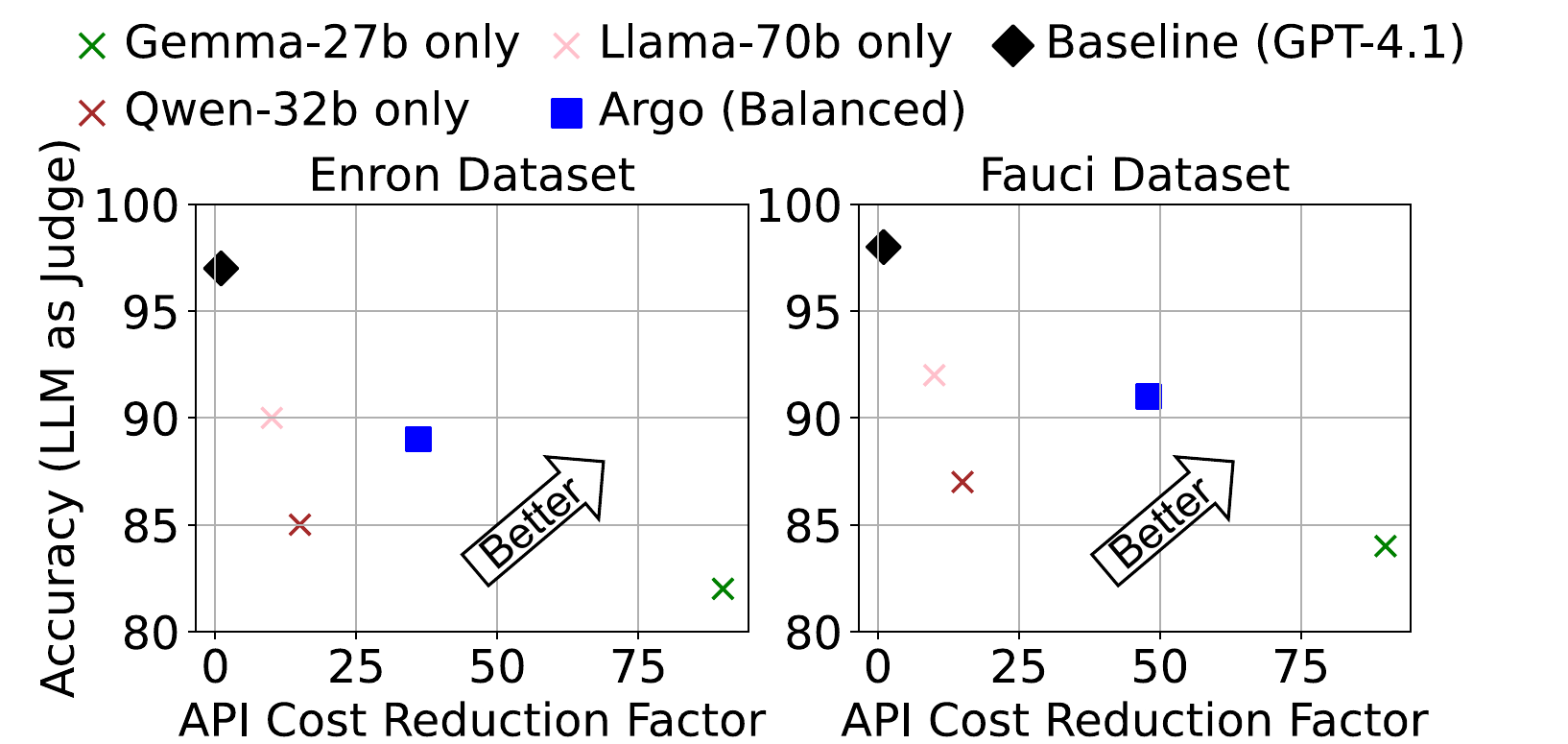}
    \tightcaption{\name extends to summarizing email content in addition to importance labeling}
    \label{fig:sum}
\end{figure}

\mypara{Including operator enforced constraints}
\name allows specifying operator constraints to the profiler, which services like \anonemail may require for data sharing restrictions and business agreements. We showcase \name under enforced constraints in Figure \ref{fig:constrained} applied on the SLM cascade. We enforce two constraints (a) on the cascade size to use a maximum of 4 SLMs and (b) not use any SLMs from the Llama family~\cite{meta_llama_3_1_8b_instruct, meta_llama_3_3_70b_instruct}. We replace models in the original SLM cascade with equivalent-cost models (comparable parameter count), namely \emph{mistralai/Mistral-7B-Instruct-v0.2}~\cite{mistral_7b_instruct} and \emph{DeepSeek Distilled}~\cite{deepseek_distilled}. \name finds a \emph{Balanced} operating point after enforcing the constraints and still achieves 122-148$\times$ cost reduction at acceptable quality for \anonemail.

\myparaq{How does fine-tuning compare against \name} We also measure the potential benefits of fine-tuning SLMs for achieving better quality-cost tradeoffs. In Figure \ref{fig:ft}, we incrementally finetune two SLMs in the cascade (\emph{meta-llama/Llama-3.1-8B-Instruct} and \emph{google/gemma-3-27b-it}) using LoRA~\cite{hu2021lora} on the Enron Dataset. We show that purely fine-tuning SLM doesn't achieve better quality or cost reduction compared to \name. Instead, \name is designed to be complementary to fine-tuning. Fine-tuned SLMs can be added to the profiling space of \name and with minimal fine-tuning, still achieves to 2-5\%$\times$ higher quality compared to using \name without it.

\mypara{Extending \name from email labeling to summarization} In Figure \ref{fig:sum}, we show how \name is minimally extended to perform email summarization. While the drop in quality is slightly higher than numeric labeling, \name still achieves 36-48$\times$ cost reduction as compared to frontier GPT4.1 within acceptable quality. As mentioned in \S \ref{sec:enhancements}, \name is extended to email summarization with minimal changes only in the metrics of evaluation (LLM as a Judge with GPT4.1) and use of token-frequency weighted confidence scores.

\tightsection{Conclusion}
\label{sec:conclusion}

We present \name, a system which enables cost-efficient email-labeling at enterprise scale, by using efficient profiling and inference techniques, alongside intelligent resource provisioning to minimize costs at peak load. Over 3 open-source email datasets, \name achieves 148-167$\times$ inference cost reduction with negligible quality degradation and 20-640000$\times$ cheaper profiling costs without any quality loss.

% %-------------------------------------------------------------------------------
\bibliographystyle{plain}
\bibliography{main}
% \bibliography{\jobname}
% \clearpage

\pagebreak
\appendix
\end{document}